\documentclass[aps,pra,twocolumn,showpacs,superscriptaddress]{revtex4}
\usepackage{graphicx}
\usepackage{latexsym}
\usepackage{amssymb}
\usepackage{color}
\usepackage{bm}
\usepackage{amsmath}
\usepackage{subfigure}
\begin{document}

\title{Heavy fermions in an optical lattice}

\author{Michael Foss-Feig}
\affiliation{
Department of Physics, University of Colorado, Boulder, Colorado 80309, USA}
\affiliation{ 
JILA, Boulder, Colorado 80309, USA}
\author{Michael Hermele}
\affiliation{
Department of Physics, University of Colorado, Boulder, Colorado 80309, USA}
\author{Victor Gurarie}
\affiliation{
Department of Physics, University of Colorado, Boulder, Colorado 80309, USA}
\author{Ana Maria Rey}
\affiliation{
Department of Physics, University of Colorado, Boulder, Colorado 80309, USA}
\affiliation{ 
JILA, Boulder, Colorado 80309, USA}
\affiliation{NIST, Boulder, Colorado 80309, USA}

\begin{abstract}
We employ a mean-field theory to study ground-state properties and transport of a two-dimensional gas of ultracold alklaline-earth metal
atoms governed by the Kondo Lattice Hamiltonian plus a parabolic confining potential.  In a homogenous system this mean-field theory is believed to give a qualitatively correct description of heavy fermion metals and Kondo insulators: it reproduces the Kondo-like scaling of the quasiparticle mass in the former, and the same scaling of the
excitation gap in the latter.  In order to understand ground-state properties in a trap we extend this mean-field theory via local-density approximation. We find that the Kondo insulator gap manifests as a shell structure in the trapped density profile.  In addition, a strong signature of the large Fermi surface expected for heavy fermion systems survives the confinement, and could be probed in time-of-flight experiments.  From a full self-consistent diagonalization of the mean-field theory we are able to study dynamics in the trap.  We find that the mass enhancement of quasiparticle excitations in the heavy Fermi liquid phase manifests as slowing of the dipole oscillations that result from a sudden displacement of the trap center.
\end{abstract}
\date{\today}
\pacs{67.85.--d, 03.75.Ss, 37.10.Jk, 71.27.+a}

\maketitle

\setlength{\parskip}{0pt plus 0.1 pt}
\setlength{\abovecaptionskip}{1pt}
\setlength{\belowcaptionskip}{0pt}

\section{Introduction}

Ultracold atoms in optical lattices offer a clean and controllable setting for the experimental investigation of condensed-matter Hamiltonians. 
A number of remarkable experiments along these lines\cite{Lewenstein,BlochRMP}, including the observation of the Mott insulator to superfluid transition of the Bose-Hubbard model using bosonic alkali-metal atoms\cite{Greiner}, and more recently the metal to insulator transition of the Fermi-Hubbard model using fermionic alkali-metal atoms \cite{schneider,jordens}, have thoroughly established the importance and feasibility of such optical lattice emulations.  Recently it has been proposed that several unique properties of fermionic alkaline-earth-metal atoms (AEMAs) make them particularly well suited for the simulation of a broad class of condensed matter Hamiltonians describing the interplay between internal (spin) and external (orbital) electronic degrees of freedom \cite{Gorshkov:2009p4747,Hermele,cazalilla, xu,fossfeig}.  These two-electron atoms are more complicated than their one-electron alkali-metal counterparts, however in the last few years a great deal of progress has been made in cooling a variety of bosonic and fermionic isotopes to quantum degeneracy \cite{fukuhara1,fukuhara2,stellmer,kraft,escobar,taie,desalvo}.  Motivated by these developments, here we discuss cold-atom probes of the Kondo Lattice Model (KLM).

\begin{figure}[h]
\subfiguretopcaptrue
\centering
\subfigure[][]{
\label{KLMgraphic}
\includegraphics[width=5.1 cm]{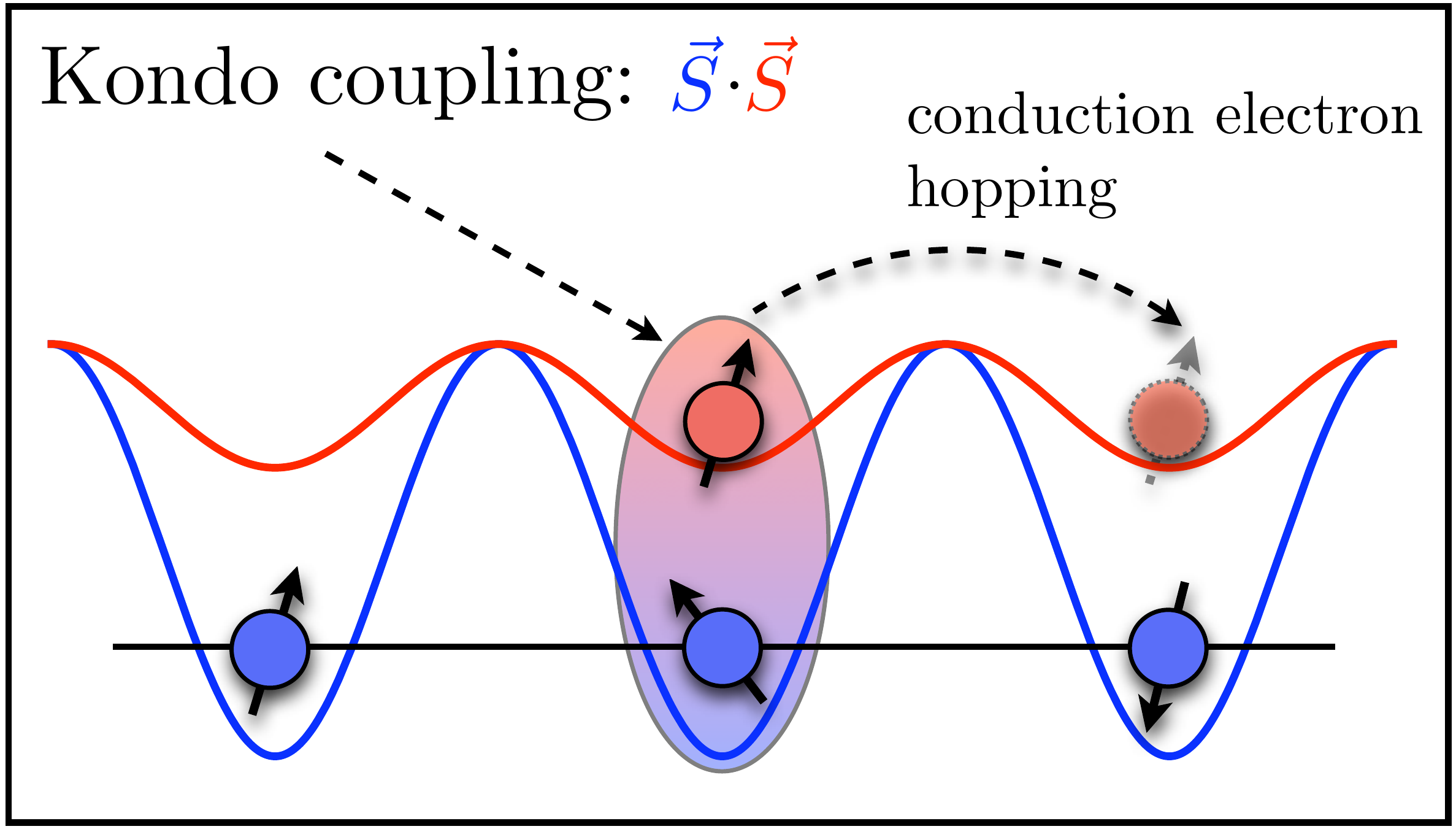}
}
\subfigure[][]{
\label{PD2D}
\includegraphics[width=3.16 cm]{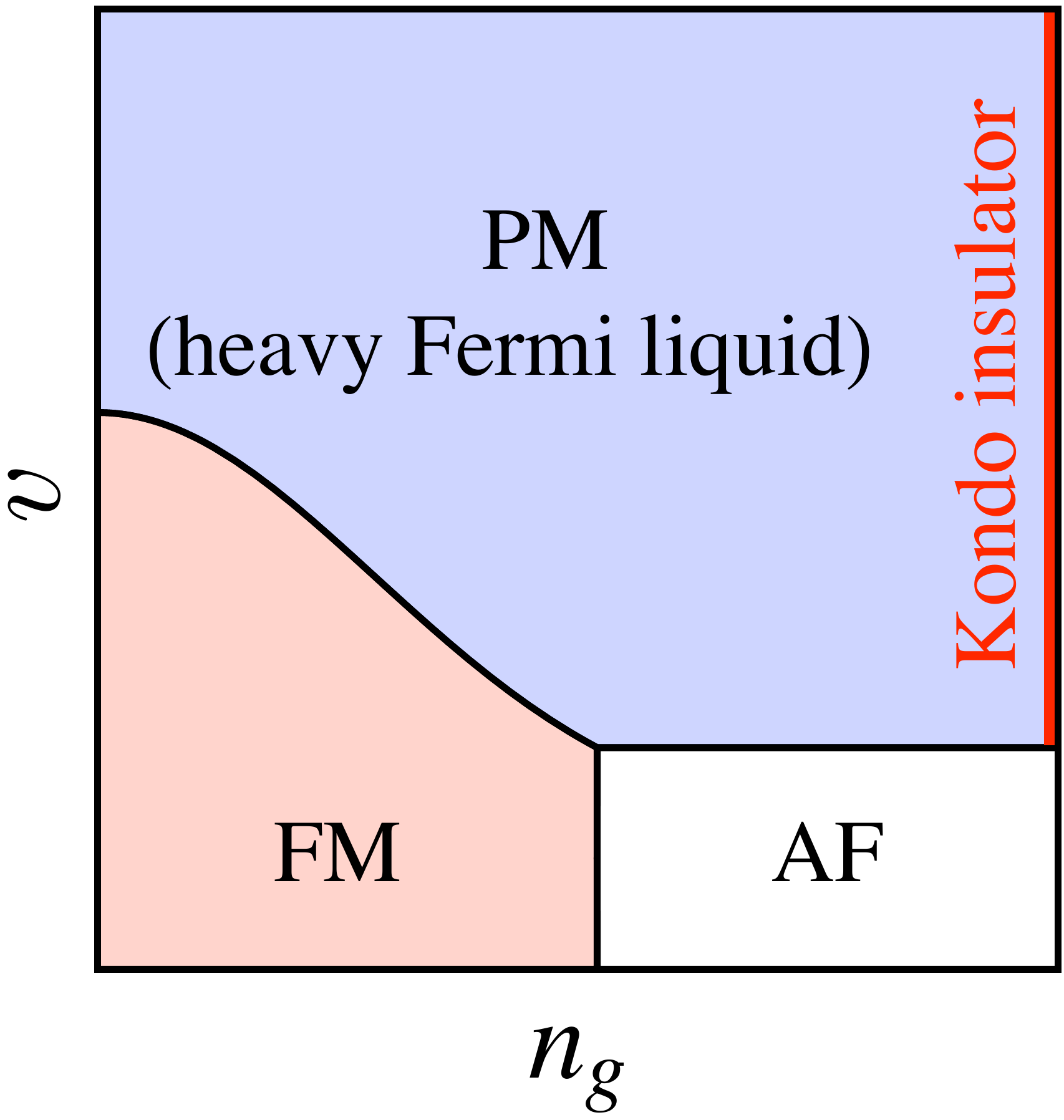}
}
\caption{\subref{KLMgraphic} In the Kondo Lattice Model the conduction electrons (red) can hop from site to site, and they interact with localized spins (blue) via a Heisenberg exchange. \subref{PD2D} 2D mean-field ground-state phase diagram as constructed in \cite{Lacroix:1979p981}, $n_g$ being conduction electron density and $v$ being a dimensionless measure of the interaction strength. PM is a paramagnetic phase, in which the heavy Fermi liquid behavior is expected.  The FM (AFM) phase is where RKKY interactions \cite{randk,kasuya,yosida} generate ferromagnetic (antiferromagnetic) order among the localized spins.  Exactly at $n_g=1$ the AF phase gives way to a non-magnetic insulating state for sufficiently large $v$ \cite{assaad}.}
\label{phasediagrams}
\end{figure}

The KLM is a canonical model for studying the interaction of conduction (mobile) electrons with localized (immobile) spin-$1/2$ scattering centers \cite{Doniach:1977}.  In the most common presentation of the model there is one localized spin on each lattice site, and the only interaction considered is an on-site Heisenberg exchange between the conduction electrons and the localized spins [see Fig. \ref{KLMgraphic}]. In a recent paper \cite{fossfeig} the implications of Ref. \cite{Gorshkov:2009p4747} for simulating the KLM in one dimension (1D) were discussed.  In particular, we suggested realistic experimental probes of physics occurring in several different regions of the 1D phase diagram.  Part of the analysis in \cite{fossfeig} showed via a mean-field calculation that important properties of the KLM's paramagnetic phase are manifest in dipole oscillations of the cold-atom system upon displacement of a confining potential.  Here we present a detailed extension of these mean-field calculations to a two-dimensional (2D) system, where the model is more closely related to real solid-state systems. 

For different values of the exchange coupling the KLM gives rise to very different physics.  We focus exclusively on the case of antiferromagnetic exchange (favoring anti-alignment of the localized spin and a conduction electron on the same site), which has been studied for over three decades as a model for inter-metallic compounds with anomalously massive quasiparticle excitations, termed \emph{heavy fermion materials} \cite{Doniach:1977,stewart,pcoleman}.  Despite the wealth of theoretical machinery developed for and brought to bear on the antiferromagnetic KLM, the structure of the ground-state phase diagram (in $D>1$) is not yet well understood.  Nevertheless, the following qualitative picture has emerged [see Fig. \ref{PD2D}].  At sufficiently weak coupling the conduction electrons mediate a long-range magnetic interaction (RKKY) between the localized spins \cite{randk,kasuya,yosida}.  The system will have magnetic order with details depending on the lattice structure and the conduction electron density.  At stronger coupling, the conduction electrons screen the localized spins by binding to them in singlets.  The system is paramagnetic, and the excitations are thought to belong to a heavy Fermi liquid (HFL).

A major outstanding challenge in KLM research is to understand the quantum phase transition that separates these two regimes, and this challenge could in principle be addressed by careful analysis of an optical lattice emulator.  But before there can be any prospect for the exploration of unknown aspects of the phase diagram, it is imperative that we understand the manifestation of relatively well understood KLM physics in an inhomogeneous setting---the ubiquitous cold-atom trap.  With this goal in mind, we present an extension to the trap of a well known mean-field theory (MFT) appropriate for describing the HFL \cite{Lacroix:1979p981}.  Some ground-state properties can be determined with relatively little effort within local-density approximation (LDA).  Already at this level important Kondo physics can be distinguished in the trap.  For example, the Kondo insulator phase induces a shell structure in the trapped density distribution, and the emergence of heavy quasiparticles shifts conduction atom weight out to the so called ``large Fermi surface''.  Outside of the range of validity of the LDA, for example in non-equilibrium situations, a numerical self-consistent diagonalization of the MFT is employed.  Using the exact mean-field wave function we explore center-of-mass oscillations resulting from a sudden displacement of the trap center, and find a clear signature of the heavy fermion mass enhancement.

The paper is organized as follows.  In Section \ref{soklm} we briefly review technical aspects of simulating the KLM with AEMAs.  In Section \ref{sec3} we introduce the MFT and review its relevance to heavy fermion physics in a translationally invariant system.  Section \ref{secLDA} exploits the translationally invariant theory to characterize important ground-state properties of the trapped system within the LDA, including density and quasi-momentum distributions.  In Section \ref{secSCD} we explain how to implement self-consistent diagonalization of the inhomogeneous MFT, and in Section \ref{secDynamics} we apply these solutions to calculations of non-equilibrium dynamics.  Section \ref{secECO} concludes with experimental considerations.

\section{Simulation of the KLM with Alkaline-Earth-Metal Atoms}
\label{soklm}

There are two primary features of AEMAs that make them suitable for simulation of the KLM: (1) The long-lived $^1S_0$ ($g$) and $^3P_0$ ($e$) clock states can be trapped independently by two different but spatially commensurate optical lattices \cite{Daley:2008p5398}.  This enables us to consider localizing one clock state in a deep lattice while keeping the other state mobile in a shallow lattice. (2) Both clock states have total electronic angular momentum $J=0$.  As a result there is essentially no coupling between the different nuclear spin states and therefore no spin changing collisions (in contrast to alkali metal atoms).  This enables us to consider an ensemble where just two states in the hyperfine manifold are populated, since to a very good approximation no transfer of states into the rest of the manifold can occur. We will identify these two populated levels with spin up and spin down of the electrons in the KLM.

In order to make the above discussion more quantitative, we must introduce some notation of Ref. \cite{Gorshkov:2009p4747} to describe and compare various energy scales.  If the lattice containing the atomic state $\alpha\in\left\{g,e\right\}$ has a potential $V_{\alpha}$ and lowest Wannier orbital $w_{\alpha}$, then we define hopping energies $J_{\alpha}=\int d^D\textbf{r}w_{\alpha}\left(\textbf{r}_{i}-\textbf{r}\right)
[-\frac{\hbar^2}{2m}\nabla^2+V_{\alpha}]w_{\alpha}\left(\textbf{r}_j-\textbf{r}\right)$ ($m$ is the atomic mass, and $(\textbf{r}_i,\textbf{r}_j)$ are the centers of two nearest neighbor sites).  In addition to these hopping energies there are a variety of possible interactions, which at sufficiently low temperatures can be parameterized in terms of the 4 $s$-wave scattering lengths $a_{ee}$, $a_{gg}$ and $a_{eg}^{\pm}$, for scattering in the states $\left|ee\right>$, $\left|gg\right>$, and $\frac{1}{\sqrt{2}}\left(\left|eg\right>\pm\left|ge\right>\right)$ respectively.  If we neglect interactions between atoms that are not on the same site, the four possible interaction energies are: $U_{\alpha\alpha}=\left(4\pi\hbar^2a_{\alpha\alpha}/m\right)\int d^D\textbf{r}w_{\alpha}^4\left(\textbf{r}\right)$ and $U_{eg}^{\pm}=\left(4\pi\hbar^2a_{eg}^{\pm}/m\right)\int d^D\textbf{r}w_{e}^2\left(\textbf{r}\right)w_{g}^2\left(\textbf{r}\right)$.

We are interested in the case where $U_{ee}\gg J_{e}$ and $U_{gg}\ll J_g$.  Under these conditions, and with a sufficiently weak confining potential $\Omega\left(i\right)$, the $e$ atoms form a unit filled Mott insulator at the center of the trap and the $g$ atoms are to a good approximation non-interacting.  We choose the $e$ atoms to represent the localized spins because they would otherwise suffer lossy collisions \cite{Gorshkov:2009p4747}.  At temperatures well below $U_{ee}$, to lowest order we can drop both $U_{ee}$ and $J_e$ and simply constrain the Hilbert space to have one $e$ atom per site.  Certainly to higher order in $J_e/U_{ee}$ there would be super-exchange interactions, but we assume these to be negligible compared to all other terms in the Hamiltonian.  Defining $V_{ex}=\left(U_{eg}^+-U_{eg}^-\right)/2$ and neglecting terms which are constant in consideration of the constraint on the $e$ atom density, what remains is an inhomogeneous version of the Kondo Lattice Hamiltonian:
\begin{eqnarray}
\label{KLM}
\mathcal{H}_{\mathrm{K}}&=&-J_g\sum_{\left<i,j\right>,\sigma}c^{\dagger}_{ig\sigma}c^{}_{jg\sigma}+V_{\mathrm{ex}}\sum_{i\sigma\sigma'}c^{\dagger}_{ig\sigma}c^{\dagger}_{ie\sigma'}c^{}_{ig\sigma'}c^{}_{ie\sigma} \nonumber \\
&+&\sum_{i}\Omega\left(i\right)\hat{n}_{ig}.
\end{eqnarray}
In Eq (\ref{KLM}) $\hat{n}_{i\alpha}=\sum_{\sigma}c^{\dagger}_{i\alpha\sigma}c^{}_{i\alpha\sigma}$, where $c^{\dagger}_{i\alpha\sigma}$ creates an atom in the lowest Wannier orbital on the site centered at $\bm{r}_i$, in electronic state $\alpha\in\{e,g\}$, and (nuclear) spin state $\sigma$.  In the rest of the paper we will consider only $D$-dimensional hypercubic lattices, with a particular focus on $D=2$.  In the first term of the Hamiltonian we have used the convention that $\langle i,j\rangle$ restricts the summation to nearest-neighbor sites.  It is important to observe that satisfaction of the inequalities $U_{ee}\gg J_e$ and $U_{gg}\ll J_g$ places no fundamental constraint on $V_{\mathrm{ex}}/J_g$.  This independence of parameters is a unique feature of the AEMA simulations proposed in Ref. \cite{Gorshkov:2009p4747}.  Previous proposals to study the KLM using alkali-metal atoms lack this tunability because they either populate multiple bands of a single lattice (Ref. \cite{duan}, in which case $U_{ee}$, $U_{gg}$ and $V_{\mathrm{ex}}$ all scale with the lattice depth), or generate Eq (\ref{KLM}) as an effective Hamiltonian which necessarily operates at weak coupling ($\left|V_{\mathrm{ex}}\right|/J_g\ll1$) \cite{paredes}.  In the present scheme, $V_{\mathrm{ex}}$ will vary from one isotope to another, and can be further adjusted by offsetting the two lattices with respect to each other (to decrease the overlap between Wannier orbitals).

\section{Mean-Field Theory}
\label{sec3}

There is no nontrivial parameter regime in which Eq. (\ref{KLM}) can be solved exactly, and approximations have to be made.  In the HFL phase a qualitative understanding can be gained by treating the interaction term at mean-field level \cite{Lacroix:1979p981}:
\begin{eqnarray}
V_{\mathrm{ex}}c^{\dagger}_{ig\sigma}c^{\dagger}_{ie\sigma'}c^{}_{ig\sigma'}c^{}_{ie\sigma}&\rightarrow&
V_{\mathrm{ex}}\sum_{\sigma'}\tilde{V}_i\left(c^{\dagger}_{ig\sigma'}c^{}_{ie\sigma'}+c^{\dagger}_{ie\sigma'}c^{}_{ig\sigma'}\right)
\nonumber \\
-V_{\mathrm{ex}}\tilde{V}_i^2&&
\end{eqnarray}
where we have defined the on-site hybridizations
\begin{equation}
\label{SelfConsistency}
\tilde{V}_i=\frac{1}{2}\sum_{\sigma}\left<c^{\dagger}_{ie\sigma}c^{}_{ig\sigma}+c^{\dagger}_{ig\sigma}c^{}_{ie\sigma}\right>.
\end{equation}
This expectation value, along with all others in this paper, is taken in the ground state of the mean-field Hamiltonian [see Eq. (\ref{HMFT}) below].  Because the mean-field Hamiltonian is quadratic, the ground state is a Slater determinant of the lowest $(N_e+N_g)/2$ single particle states (half the total number of atoms due to spin degeneracy), where $N_{\alpha}$ is the total number of $\alpha$ atoms.  This decoupling is not unique, for instance there could be terms that mix different spin states, but if the KLM is generalized to allow the spin index $\sigma$ to run from $1$ to $N$ ($N=2$ for the KLM), then the above decoupling is exact in the limit $N\rightarrow\infty$ \cite{Coleman:2009p2107,Read:1984p2396}.  One should keep in mind, however, that the decision not to keep any spin mixing terms in the decoupling guarantees that this MFT cannot capture a transition to a magnetically ordered state, and therefore cannot remain valid for arbitrarily small $v$ (where RKKY-induced magnetic order should exist).  The resulting quadratic Hamiltonian is
\begin{align}
\label{HMFT}
\mathcal{H}_{\mathrm{M}}=-J_g\sum_{\left<i,j\right>,\sigma}c^{\dagger}_{ig\sigma}c^{}_{jg\sigma}+\sum_{i}\left[\Omega\left(i\right)\hat{n}_{ig}+\mu_{ie}\left(\hat{n}_{ie}-1\right)\right] \nonumber \\+V_{\mathrm{ex}}\sum_{i\sigma}\tilde{V}_i\left(c^{\dagger}_{ig\sigma}c^{}_{ie\sigma}+c^{\dagger}_{ie\sigma}c^{}_{ig\sigma}\right)-V_{\mathrm{ex}}\sum_{i}\tilde{V}_i^2,
\end{align}
where the $\mu_{ie}$ are local chemical-potentials needed to enforce the constraint of one $e$ atom per site on average (in the exact KLM we have the much stronger constraint $\langle(\hat{n}_{ie}-1)^2\rangle=0$, but in the MFT this must be relaxed to $n_{ie}\equiv\langle \hat{n}_{ie}\rangle=1$).  The theory is now paramagnetic, because $\mathcal{H}_{\mathrm{M}}$ does not couple different spin states, and the eigenstates are created by quasiparticle operators:
\begin{equation}
\label{BMCs}
\alpha^{\dagger}_{q\sigma}=\sum_{i}\left(u^{i}_qc^{\dagger}_{ig\sigma}+v^{i}_qc^{\dagger}_{ie\sigma}\right).
\end{equation}
Here $q\in\{1,2,...\}$ is an index that labels the different quasiparticle eigenstates, and not a wave vector.  Solving the MFT then means finding the above mode coefficients $u^i_q$ and $v^i_q$, but this needs to be done self-consistently: As parameters in the Hamiltonian the $\tilde{V_i}$ determine the mode coefficients, but in turn the mode coefficients determine the $\tilde{V}_i$ via the definition
\begin{eqnarray}
\label{vref}
\tilde{V}_i&=&\frac{1}{2}\sum_{\sigma}\left<c^{\dagger}_{ie\sigma}c^{}_{ig\sigma}+c^{\dagger}_{ig\sigma}c^{}_{ie\sigma}\right> \nonumber \\
&=&\sum_{q=1}^{(N_g+N_e)/2}\left(u^i_q\bar{v}^i_q+v^i_q\bar{u}^i_q\right).
\end{eqnarray}
We sum over a number of modes equal to half of the total number of particles, but put two particles in each mode to account for the degeneracy of spin.  Similarly, the onsite $g$ and $e$ atom densities are given by
\begin{eqnarray}
n_{ig}=\sum_{\sigma}\left<c^{\dagger}_{ig\sigma}c^{}_{ig\sigma}\right>=2\sum_{q=1}^{(N_g+N_e)/2}u^i_q\bar{u}^i_q \nonumber \\
n_{ie}=\sum_{\sigma}\left<c^{\dagger}_{ie\sigma}c^{}_{ie\sigma}\right>=2\sum_{q=1}^{(N_g+N_e)/2}v^i_q\bar{v}^i_q.
\end{eqnarray}
If we define the ground-state energy $E=\left<\mathcal{H}_{\mathrm{M}}\right>$, the self-consistency condition and the local constraints $n_{ie}=1$ can be written compactly as
\begin{equation}
\partial E/\partial\tilde{V}_i=0~~\mathrm{and}~~\partial E/\partial\mu_{ie}=0
\end{equation}
respectively.

\subsection{Translationally Invariant Case}
Before solving the MFT in the presence of a trap, it is useful to review known results for the translationally invariant case ($\Omega=0$) \cite{Lacroix:1979p981}.  It is common to assume that the hybridizations and chemical potentials retain the discrete translational invariance of the exact KLM Hamiltonian  ($\tilde{V}_i,\mu_{ie}\rightarrow \tilde{V},\mu_e$), in which case the Hamiltonian in Eq. (\ref{HMFT}) reduces to
\begin{eqnarray}
\label{HMFTt}
\mathcal{H}_{\mathrm{T}}&=&-J_g\sum_{\left<i,j\right>\sigma}c^{\dagger}_{ig\sigma}c^{}_{jg\sigma}+\mu_e\left(\sum_{i}\hat{n}_{ie}-\mathcal{N}\right) \nonumber \\
&+&V_{\mathrm{ex}}\sum_{i\sigma}\tilde{V}\left(c^{\dagger}_{ig\sigma}c^{}_{ie\sigma}+c^{\dagger}_{ie\sigma}c^{}_{ig\sigma}\right)-\mathcal{N}V_{\mathrm{ex}}\tilde{V}^2,
\end{eqnarray}
$\mathcal{N}$ being the total number of lattice sites.  It is common to include a $g$ atom chemical potential by adding a term $-\mu_g\sum_{i}\hat{n}_{ig}$ in Eq. \ref{HMFTt}, but we find it conceptually simpler to work at a fixed $n_g$ for the time being (we will trade in $n_g$ for $\mu_g$ in the next section to apply the LDA).  Due to the translational invariance the nonconstant part of $\mathcal{H}_{\mathrm{T}}$ (those terms containing operators) can easily be diagonalized for general $\tilde{V}$ and $\mu_e$ by going to Fourier space.  Defining $c_{\bm{k}\alpha\sigma}=\frac{1}{\mathcal{\sqrt{N}}}\sum_{j}c_{j\alpha\sigma}e^{i\bm{r}_j\cdot\bm{k}}$ (where $\bm{r}_j$ is the position of site $j$ and $\bm{k}$ is a quasi-momentum vector in the first Brillouin zone), we find quasiparticles
\begin{equation}
\alpha^{\dagger}_{\bm{k}\sigma}=c^{\dagger}_{\bm{k}g\sigma}u(\bm{k})+c^{\dagger}_{\bm{k}e\sigma}v(\bm{k})
\end{equation}
and the quasiparticle spectrum
\begin{equation}
\label{mftdispersion}
\mathcal{E}_{\pm}(\bm{k},\tilde{V},\mu_e)=\frac{\epsilon\left(\bm{k}\right)+\mu_e}{2}\pm\frac{1}{2}\sqrt{4V_{\mathrm{ex}}^2\tilde{V}^2+\left(\epsilon\left(\bm{k}\right)-\mu_e\right)^2}.
\end{equation}
Here $\epsilon\left(\bm{k}\right)=-2J_g\sum_{i=1}^D\cos(k_ia)$ ($a$ being the lattice spacing) is the tight-binding dispersion for the $g$ atoms in a $D$-dimensional hyper-cubic lattice, and the two branches of the quasiparticle spectrum are called the upper and lower hybridized bands.  If $\mathcal{N}$ is large then the energy per site of the translationally invariant system, given by $E(\tilde{V},\mu_e,n_g)=\left<\mathcal{H}_{\mathrm{T}}\right>/\mathcal{N}$, can be written as an integral over the Fermi volume $\mathcal{F}$.  For example, on a $D$ dimensional lattice with $n_g<1$ the total energy per site is:
\begin{equation}
E(\tilde{V},\mu_e,n_g)=-V_{\mathrm{ex}}\tilde{V}^2-\mu_e+\frac{1}{(2\pi)^D}\int_{\mathcal{F}}d^Dk \ \mathcal{E}_-(\bm{k},\tilde{V},\mu_e),
\label{RE}
\end{equation}
with the dependence on $n_g$ hidden in the limits of integration.  Only the lower hybridized band is populated because $n_g=1$ corresponds to completely filling the first Brillouin zone (since there is also one localized atom per site).  For $1<n_g<2$ the lower hybridized band will be completely full and the upper band will be partially integrated over.  For a given $n_g$, the correct $\mu_e$ and self consistent $\tilde{V}$ can be found by solving the coupled equations
\begin{equation}
\label{scc}
\frac{\partial E(\tilde{V},\mu_e,n_g)}{\partial\tilde{V}}=0\ \ \mathrm{and} \ \ \frac{\partial E(\tilde{V},\mu_e,n_g)}{\partial\mu_e}=0.
\end{equation}
It will be useful for later application of the LDA to appreciate that the only knob we can turn to determine the MFT solutions (beyond the couplings $J_g$ and $V_{\mathrm{ex}}$ in the Hamiltonian) is a choice of the $g$ atom density.  Once $n_g$ is chosen, the correct $\tilde{V}$ and $\mu_e$ are fixed by Eq (\ref{scc}).

\begin{figure}[h]
\centering
\subfiguretopcaptrue
\subfigure[][]{
\includegraphics[width=6 cm]{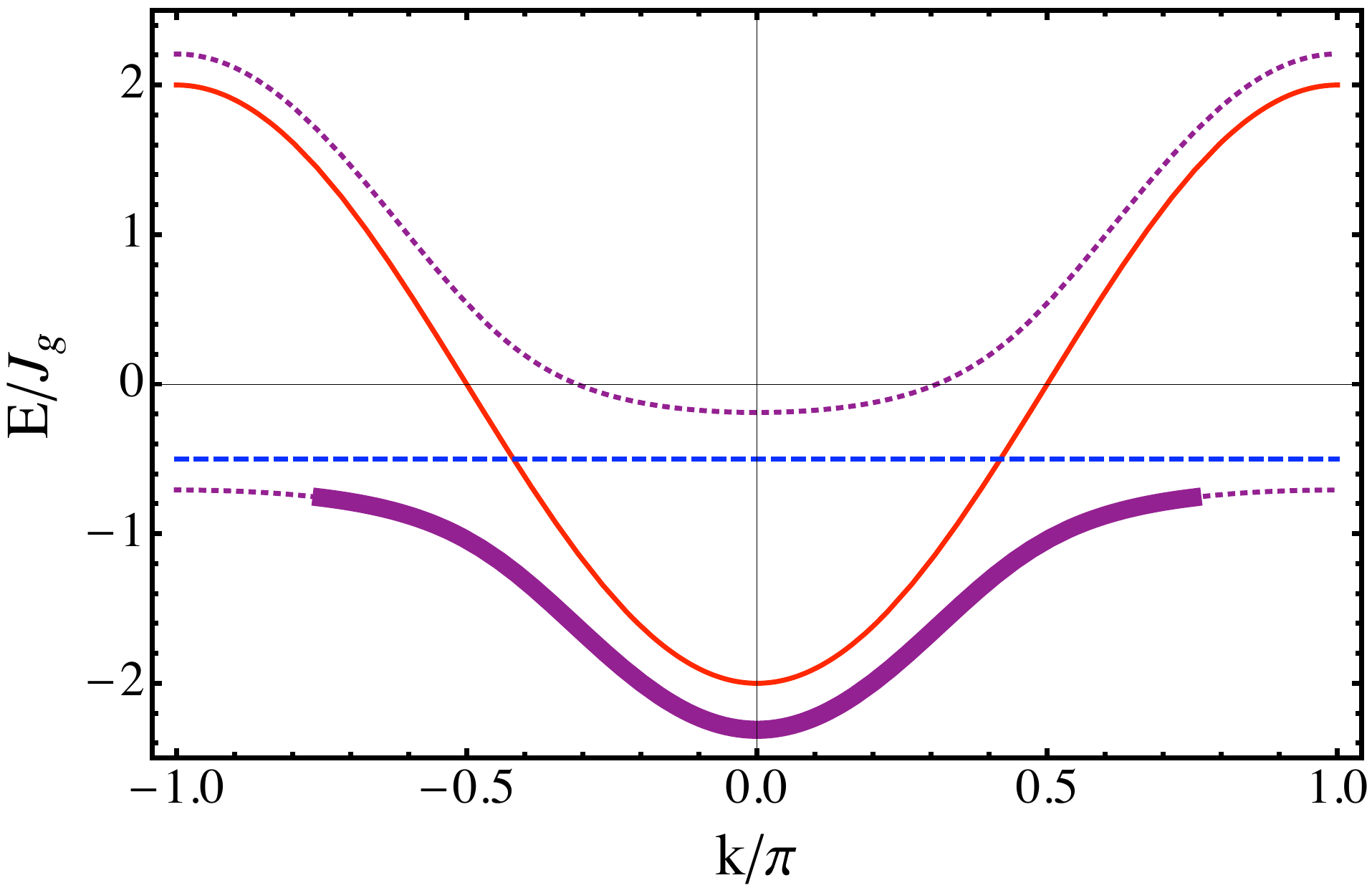}
\label{phyb2}
}
\subfigure[][]{
\includegraphics[width=6 cm]{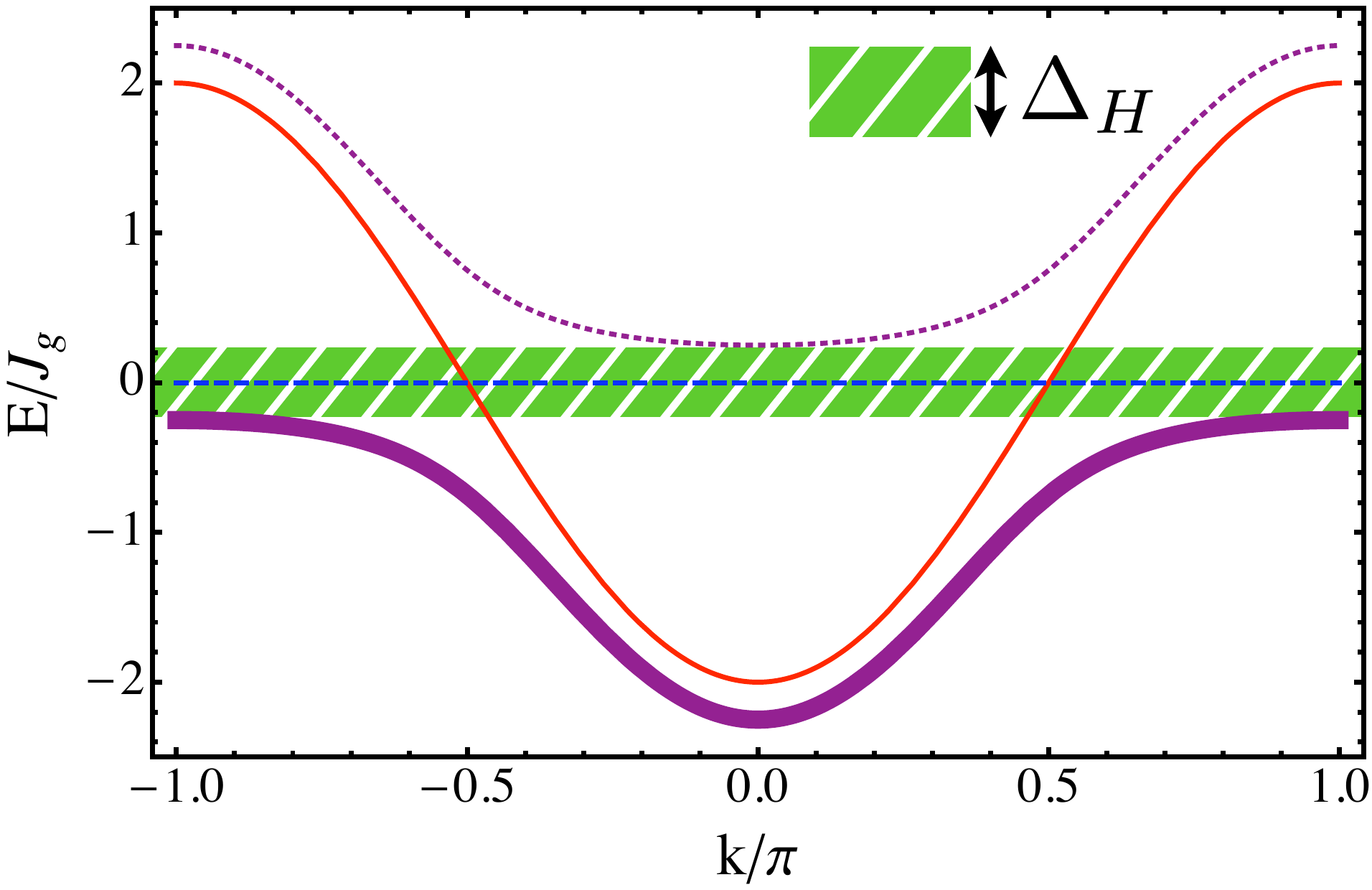}
\label{phyb3}
}
\caption{Spectrum of the translationally invariant MFT for $n_g<1$ \subref{phyb2} and for $n_g=1$ \subref{phyb3}, shown in 1D for clarity.  The blue dashed line is the $e$ atom band ($E=\mu_e$), and the red solid curve is the $g$ atom band ($E=-2J_g\cos(k)$).  The purple dotted lines are the hybridized bands $\mathcal{E}_{\pm}$, with the thicker solid section covering the Fermi volume.  In \subref{phyb3} the lower band is filled and separated from the upper band by the hybridization gap $\Delta_{\mathrm{H}}$ (green slashed region), causing the system to be insulating for $n_g=1$.}
\label{phyb}
\end{figure}

Defining the dimensionless coupling
\begin{equation}
v\equiv-2V_{\mathrm{ex}}/J_g,
\end{equation}
one finds that for small $v$ the hybridization $\tilde{V}$ will also be small.  The two hybridized bands track the $e$ and $g$ atom bands closely, however mixing due to finite $\tilde{V}$ causes the crossing to be avoided [Fig. (\ref{phyb})].  For less than unit filling of the $g$ atoms, the ground state is obtained by filling the lower hybridized band with quasiparticles out to the Fermi surface ($k_F=\frac{\pi}{2}\left(1+n_g\right)$ in 1D).  The part of the lower hybridized band running nearly parallel to the line $E=\mu_e$ [blue line in Fig. \ref{phyb2}] holds most of the $e$ atom weight, which indicates that for decreasing $n_g$ we must pull $\mu_e$ down below the center of the $g$ atom band in order to satisfy $n_e=1$.  This pins the Fermi surface to the flattened part of the hybridized band (called the Kondo resonance), which is the origin of the quasiparticle mass enhancement.

When $n_g=1$ the lower hybridized band is completely full, and $\mu_e=0$ because of a particle hole symmetry \cite{Tsunetsugu:1997p597}.  The system is an insulator with a gap $\Delta_{\mathrm{H}}$ [Fig. \ref{phyb3}].  Despite having the appearance of a band gap in the MFT, this gap should be understood as arising from correlation effects: Filling the lower hybridized band means having one conduction atom per unit cell, so non-interacting band theory of the conduction atoms would predict a metallic state.  From Eqs. (\ref{HMFT}) and (\ref{vref}) one can see that at $v=\infty$ and $\mu_e=0$ we have $\tilde{V}=1$.  It then follows from Eq (\ref{mftdispersion}) that the MFT predicts $\Delta_{\mathrm{H}}\sim J_gv$ at strong coupling.  At weak coupling, more careful analysis shows that the MFT predicts a non analytic gap scaling $\Delta_{\mathrm{H}}\sim J_ge^{-1/v}$.  This weak coupling expression should be understood as correct for $v$ small but not less than some critical $v_c>0$.  The critical coupling $v_c$ marks the phase transition between the HFL and a (presumably) magnetically ordered phase, in which the MFT is not valid [see Fig. \ref{PD2D}].

It has been implicit in the discussion above that both the $e$ and $g$ atoms are to be included in the Fermi volume of the MFT ground state.  However, one should keep in mind that the actual size of the Fermi surface is a subtle and important issue in the KLM \cite{martin, Tsunetsugu:1997p597}.  Looking at Eq. (\ref{KLM}), one might expect the volume of the Fermi sea in any KLM ground state to be determined by the number of $g$ atoms, since the $e$ atoms represent localized spins with no charge degrees of freedom.  However, the KLM at small $v$ is an effective model for the Periodic Anderson Model at large $U$, for which Luttinger's theorem \cite{luttinger} is expected to hold \cite{martin} (meaning that the Fermi volume should be determined by the number of $g$ and $e$ atoms). These two scenarios are referred to respectively as a small and large Fermi surface.  The HFL ground state, which is described by our MFT, is known to have a large Fermi surface.  The Kondo insulator is a cousin of the HFL, which arises when the density of $g$ atoms increases to unity, and the large Fermi surface of the HFL completely fills the Brillouin zone.  Our focus in this paper is on the HFL and Kondo insulator, largely because we believe it is likely that the HFL and Kondo insulator phases will occupy a significant part of the paramagnetic region of the phase diagram [PM in Fig.~\ref{PD2D}].  Indeed, on the square lattice that we consider the Kondo insulator is known to occur for $v \gtrsim 1.45$ (when $n_g = 1$) \cite{assaad}.  Given the close relationship between HFL and Kondo insulator, the HFL most likely \emph{at least} exists nearby (\emph{i.e.} for $v \gtrsim 1.45$ and $n_g \lesssim 1$).

\section{Local-Density Approximation}
\label{secLDA}

The simplest way to extend the MFT to an inhomogeneous system is via the local density approximation (LDA).  The idea is that on a given site, we consider the trapping potential to be a chemical potential in a translationally invariant system with Hamiltonian
\begin{equation}
\mathcal{H}_{\mathrm{T}}-\mu_{jg}\sum_{i\sigma}c^{\dagger}_{ig\sigma}c^{}_{ig\sigma},
\end{equation}
with $\mu_{jg}=\mu_g-\Omega(j)$.  On physical grounds one might argue that the local chemical potential should be applied to the $e$ atoms as well.  However this is unnecessary; the $e$ atom density is fixed to be one per site, so including the effect of the trap on them simply provides an overall constant energy shift (more technically, if it were applied to the localized atoms as well the local $\mu_{je}$ would simply readjust to absorb it).  We now have one translationally invariant problem per lattice site, each of which can be solved as described in Section \ref{sec3}.  There is a minor complication owing to the choice in Section \ref{sec3} to work at fixed density, not fixed chemical potential.  But this can be resolved easily by remembering that at zero temperature the chemical potential is the energy required to add one quasiparticle at the Fermi surface.  Solving the translationally invariant model at fixed $\mu_{jg}$ simply means finding the $n_{jg}$ satisfying
\begin{equation}
\frac{dE(\tilde{V},\mu_e,n_{jg})}{dn_{jg}}=\mu_{jg},
\end{equation}
since the left hand side is the energy cost of adding one quasiparticle at the Fermi surface.
The derivative on the left hand side can be expanded as
\begin{equation}
\frac{\partial E}{\partial\tilde{V}}\frac{d\tilde{V}}{d n_{jg}}+\frac{\partial E}{\partial\mu_e}\frac{d\mu_e}{d n_{jg}}+\frac{\partial E}{\partial n_{jg}},
\end{equation}
and the first two terms are zero for the MFT solution by Eq. (\ref{scc}).  The last term is the change in the energy, at fixed $\tilde{V}$ and $\mu_e$, due to the addition of  a single quasiparticle at the Fermi surface, so we conclude that
\begin{equation}
\mu_{jg}=\mathcal{E}_{\pm}(\bm{k}_F(n_{jg}),\tilde{V}(n_{jg}),\mu_e(n_{jg}))
\label{ldasolve}
\end{equation} 
Here $\bm{k}_F$ is any vector falling along the Fermi surface consistent with the filling $n_{jg}$, and the $\pm$ means to choose the branch that yields a solution: there will only be one, since both branches are monotonic functions of $n_g$ and they are separated by a gap.  Solving Eq. (\ref{ldasolve}) for $n_{jg}$ on each lattice site, while adjusting $\mu_g$ to obtain the correct total number of particles, constitutes the LDA solution.  The LDA gives us immediate access to many important groundstate properties, such as the real-space and momentum-space density distributions.

\subsection{Real space $g$ atom density distribution}
\label{rsd}


\begin{figure}[h]
\centering
\includegraphics[width=6cm]{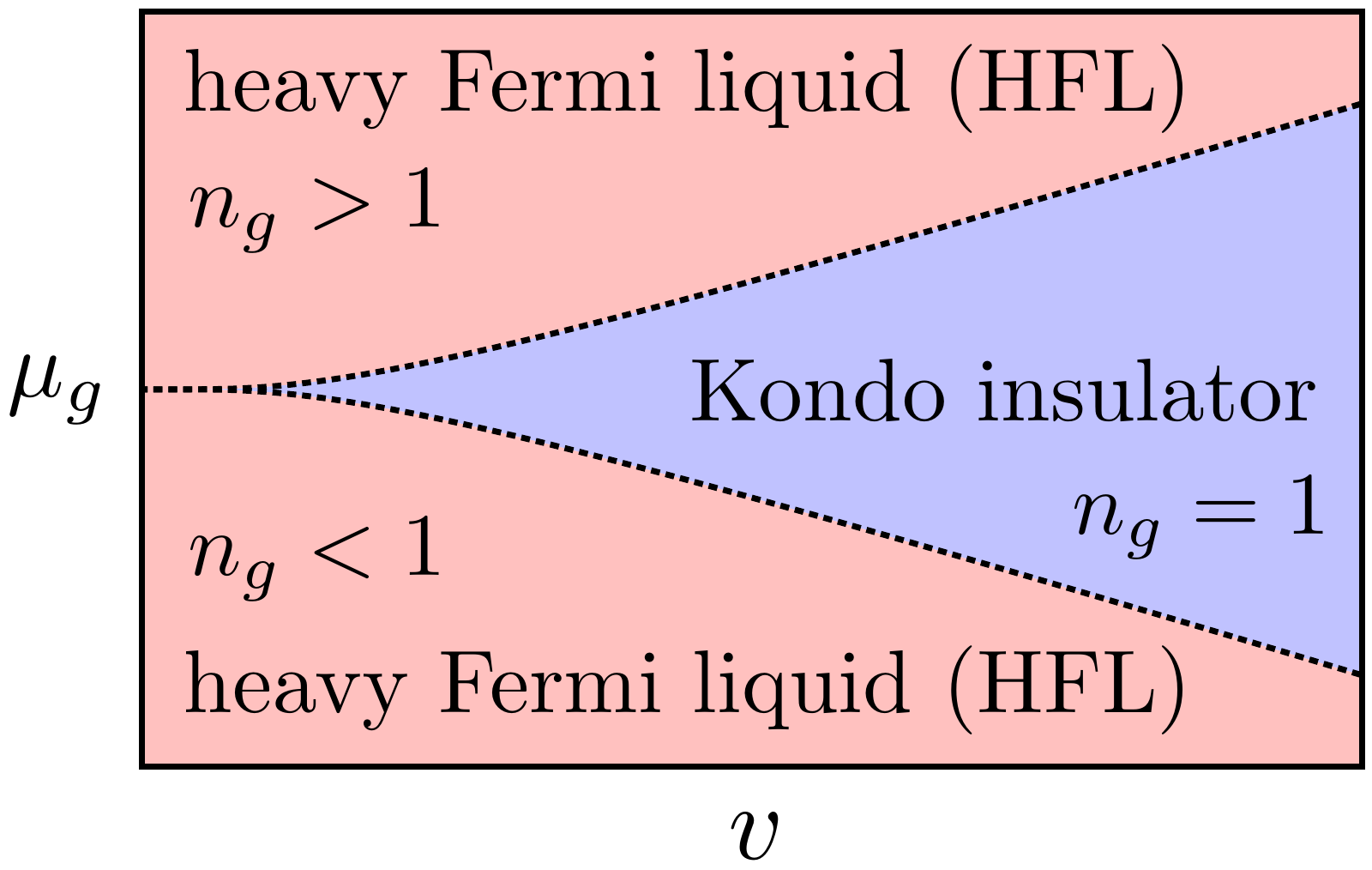}
\caption{Schematic ground-state mean-field phase diagram of the KLM as a function of chemical potential $\mu_g$ and dimensionless coupling $v$.  The black dotted lines are given by $\mu_g=\pm\Delta_{\mathrm{H}}/2$, which crosses over from Kondo like scaling ($\Delta_{\mathrm{H}}\sim J_ge^{-1/v}$) at small $v$ to linear scaling ($\Delta_{\mathrm{H}}=J_gv$) at large $v$.}
\label{mjpd}
\end{figure}

In order to understand the qualitative features of trapped $g$ atom density distributions it is helpful to plot the ground-state mean-field phase diagram in the $\mu_g-v$ plane [Fig. \ref{mjpd}].  Moving from the outside of the $g$ atom cloud towards the center of the trap corresponds to moving from the bottom to the top of the phase diagram along a line of fixed $v$.  For sufficiently small $N_g$, we will reach the trap center before entering the region of Kondo insulator, and therefore we expect to be everywhere in the $n_g<1$ heavy fermion metallic state.  If, however, we choose $N_g$ sufficiently large, we will breach the $n_g=1$ Kondo insulator phase and a unit filling plateau should develop at the center of the trap.  Increasing $N_g$ further we will find that near the center of the trap we exit the Kondo insulator phase into the $n_g>1$ heavy fermion metallic phase.  This behavior is very similar to the Mott plateaus of the Hubbard model \cite{Campbell:2006p5171,Folling:2006p5135}, only here we have a Kondo insulator and not a Mott insulator.  We also emphasize an unusual feature of the shell structure, that it exists in the density distribution of $g$ atoms, which \emph{do not interact with each other directly}.

\begin{figure}[h]
\subfiguretopcaptrue
\centering
\subfigure[][~$n_g$]{
\includegraphics[width=4 cm]{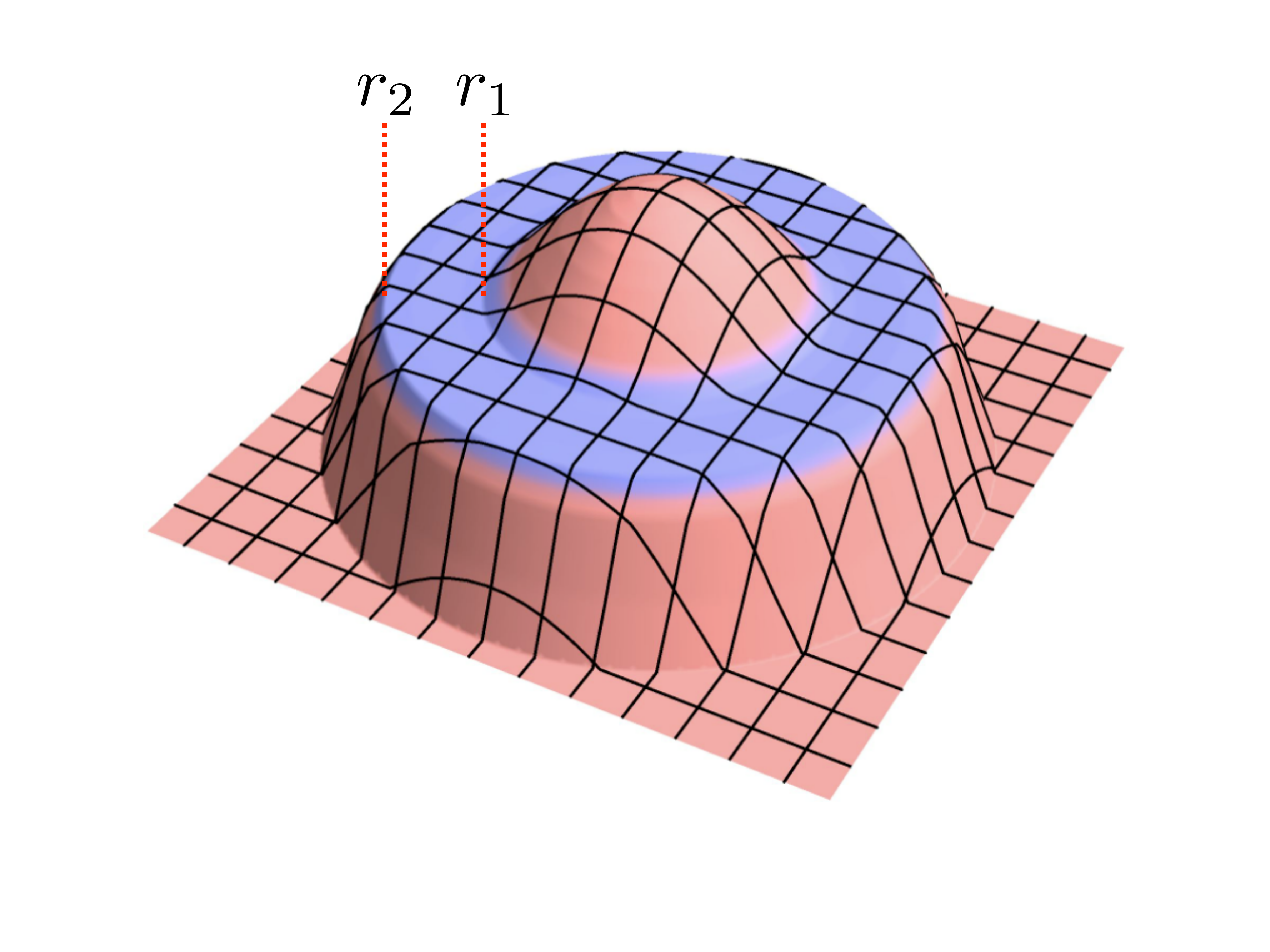}
\label{lda2da}
}
\subfigure[][~$\tilde{V}$]{
\includegraphics[width=4 cm]{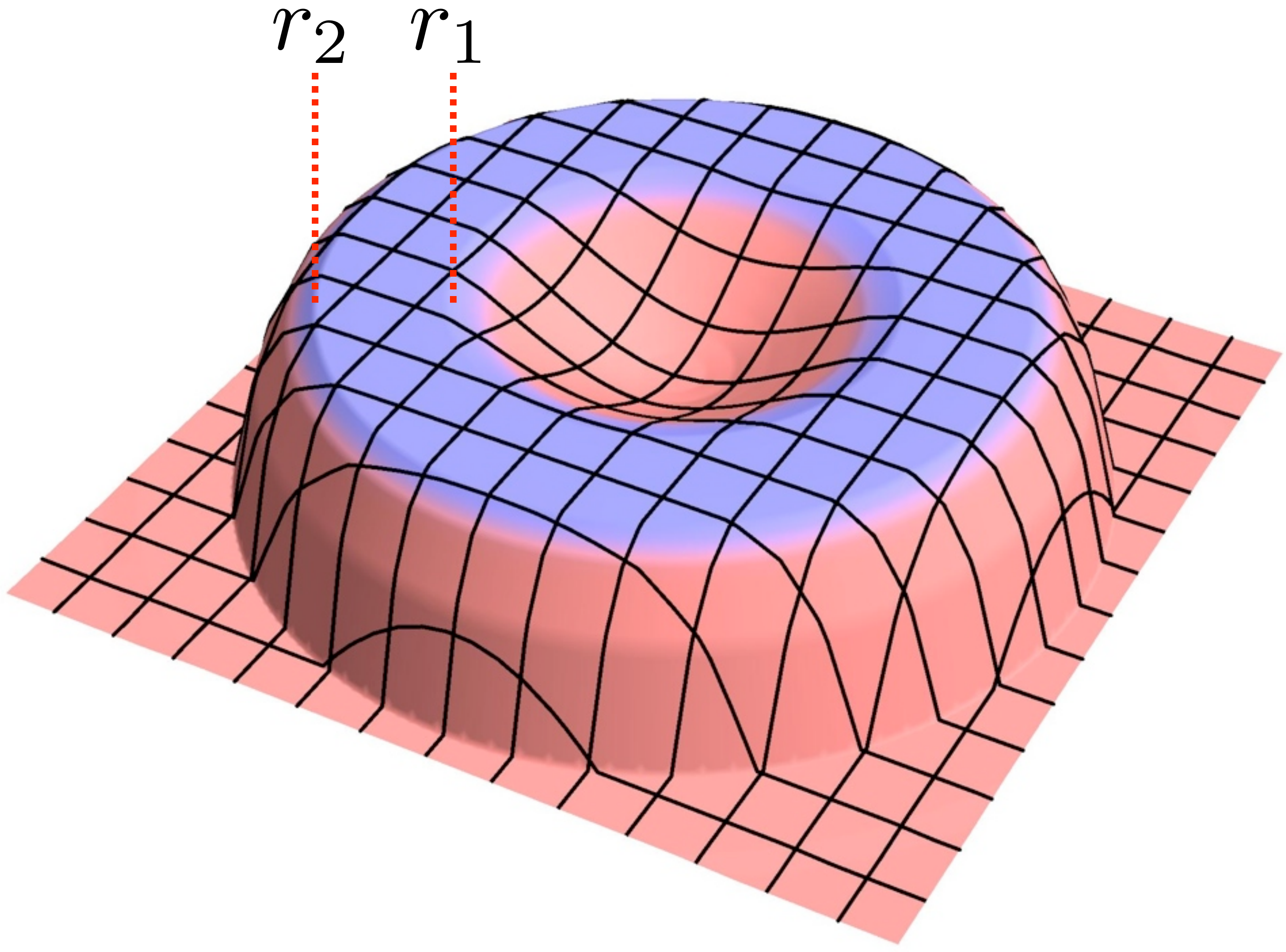}
\label{lda2db}
}
\caption{\subref{lda2da} The formation of a Kondo plateau between $r_1$ and $r_2$ at high trap fillings, obtained within LDA.  \subref{lda2db} The hybridization also displays the plateau, but as the density increases past $n_g=1$ the hybridization goes back down.  In both \subref{lda2da} and \subref{lda2db} the parameters used are $N_g\approx 550$, $\Omega/J_g= 45/1000$, and $v=8$.}
\label{lda2d}
\end{figure}

\begin{figure*}[htp]
\centering
\includegraphics[width=18cm]{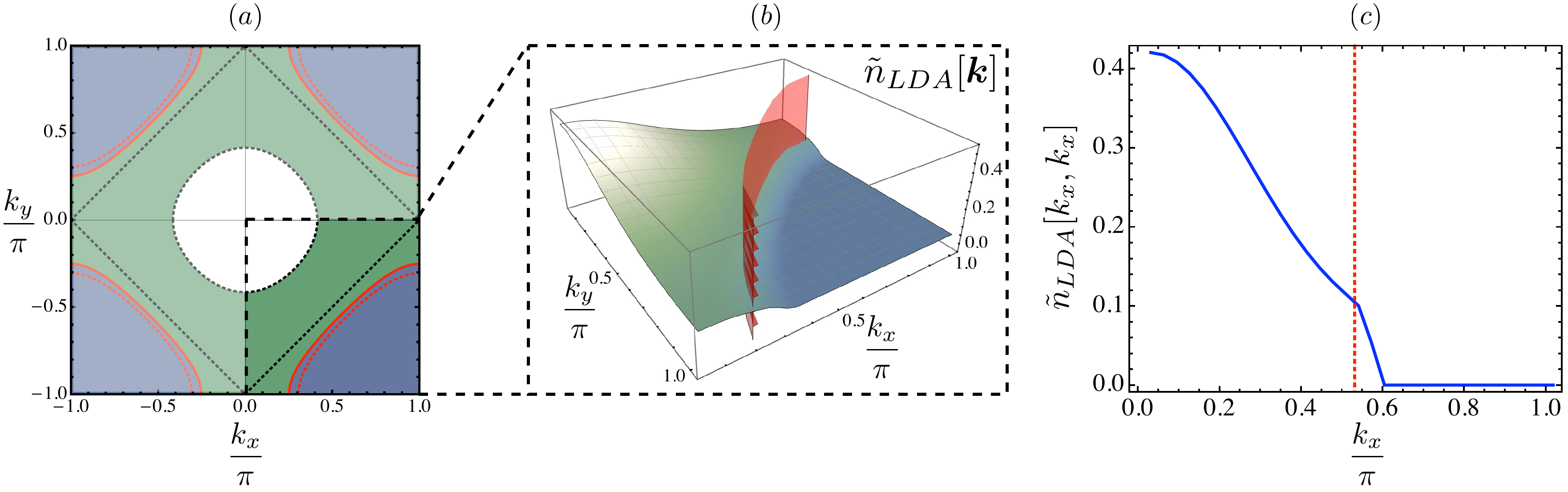}
\caption{(a) The first Brillouin zone of a 2D square lattice.  The central white region (within the dotted circle) is the small Fermi sea of the translationally invariant KLM with a conduction atom density $n_g\approx0.19$.  The shaded green region (within the solid red line) fills out the large Fermi sea  (the red solid line is the large Fermi surface, as defined in Section \ref{qmd}).  The red dotted line (just outside of the red solid line) is the large Fermi surface corresponding to a uniform filling equal to the density at the center of the trap.  Even when there are no conduction atoms the large Fermi sea occupies half of the zone (within the dotted diamond), since there is still one localized atom per site contributing to the volume. (b) Quasi-momentum distribution of the conduction atoms plotted over one quadrant of the Brillouin zone, showing a prominent feature at the large Fermi surface (red ribbon). (c) A 1D cut along the line $k_y=k_x$ in (b), $\tilde{n}_{LDA}[k_x,k_x]$ is the blue solid line, and the large Fermi surface is the red dotted line.  For all plots the parameters used are $N_g\approx116$, $\Omega/J_g=5/1000$, and $v=8$.}
\label{momentumdistributions}
\end{figure*}

In Fig. \ref{lda2da} we demonstrate such a plateau for a symmetric trap geometry $\Omega(i)=\Omega r^2$ (where $r$ is the distance of site $i$ from the origin, in lattice units).  This result was obtained by exploiting the rotational symmetry and solving a simpler 1D problem in the radial coordinate $r$.  We are still solving Eq (\ref{ldasolve}), but now instead of $\mu_{jg}$ we use $\mu_{rg}=\mu_g-\Omega r^2$, and we rotate the solution around the axis of cylindrical symmetry to obtain the plot in Fig. \ref{lda2da}.  Of course for this to work the $g$ atom cloud radius must be relatively large in lattice units, otherwise the lack of rotational symmetry of the lattice itself would be relevant.  The plateau also shows up in the hybridization profile [Fig. \ref{lda2db}], however near the center of the trap the hybridization dips back down.  Physically this occurs because for $n_g>1$ some g atoms are forced to pair with each other into singlets, reducing the number available to hybridize with the localized atoms.

Defining $r_{1\left(2\right)}$ to be the interior (exterior) radius of the plateau, it is clear from the argument used to link Fig. \ref{mjpd} to Fig. \ref{lda2d} that $\Omega\left(r_2^2-r_1^2\right)=\Delta_{\mathrm{H}}$.  More generally, in an exact treatment of the KLM the LDA solution will predict $\Omega\left(r_2^2-r_1^2\right)=2\Delta_{\mathrm{qp}}$, where $\Delta_{\mathrm{qp}}$, called the quasiparticle gap, is the difference in energy between the unit-filled ground state and the lowest energy state with one $g$ atom added or removed (these are both the same because of particle hole symmetry) \cite{Tsunetsugu:1997p597}.  Physically this result reflects the LDA assumption that transfer of density from one site to another does not affect the energy to first order ($dE(n_{jg})/d n_{jg}+\Omega(j)=\mu_g$ is constant across the trap).  Therefore the energy needed to add a quasiparticle at $r_1$ ($\Delta_{\mathrm{qp}}-\mu_g+\Omega r_1^2$) must be opposite to the energy needed to remove one at $r_2$ ($\Delta_{\mathrm{qp}}+\mu_g-\Omega r_2^2$), giving $\Omega\left(r_2^2-r_1^2\right)=2\Delta_{\mathrm{qp}}$.

At strong coupling, where the plateau is most visible, the exact quasiparticle gap of the KLM can be calculated by considering a single site (due to the relatively small hopping) \cite{Tsunetsugu:1997p597}.  If on one site we have $n_g=n_e=1$ initially, adding or subtracting a $g$ atom costs $(3/4)J_gv=\Delta_{\mathrm{qp}}$.  Therefore we can be certain that for large $v$ the plateau size should satisfy $\Omega(r_2^2-r_1^2)=(3/2)J_gv$.  In this limit $\Delta_{\mathrm{H}}=J_gv$, so the MFT underestimates the size of the plateau in the strong-coupling limit.

\subsection{$g$ atom quasi-momentum distributions}
\label{qmd}

At mea-field level a large Fermi surface arrises by assumption: By assigning a nonzero value to the hybridization matrix element $\tilde{V}$, the $e$ atoms are liberated into the Fermi sea.  This is well known, and proves nothing about the Fermi surface in the actual KLM ground state.  However, as an example of how the structure of the large Fermi surface survives in the trap, we proceed to calculate the quasi-momentum distribution in the LDA \cite{giorgini}.  In the translationally invariant MFT, there is a unique $g$ atom quasi-momentum distribution
\begin{equation}
\tilde{n}[\bm{k},n_g]=f[\bm{k},n_g]\left|u(\bm{k})\right|^2
\end{equation}
associated with every possible filling $n_g$, where $f[\bm{k},n_g]$ is the zero temperature Fermi function for non-interacting electrons at a filling $1+n_g$.  The LDA approximation to the quasi-momentum distribution is then obtained by summing $\tilde{n}$ over the densities on each lattice site
\begin{equation}
\tilde{n}_{LDA}[\bm{k}]=\sum_{j}\tilde{n}[\bm{k},n_{jg}],
\end{equation}
where the $n_{jg}$ are calculated as discussed in the previous section.

The existence of the trap introduces an unavoidable ambiguity in the definition of the small and large Fermi surfaces, since the filling is affected by what one chooses as the system size (i.e. should the sites outside of the $g$ atom cloud be included?).  However we find it reasonable, for the sake of comparison, to choose the system size to be defined by the diameter of the $g$ atom cloud.  Then the trap can be considered as a perturbation on the translationally invariant system that is sufficiently small as to not force the density to zero anywhere.  The large Fermi surface then belongs to a translationally invariant system of this size containing the same number of conduction atoms as there are in the trap.  Applying this definition to $\sim116$ conduction atoms in the symmetric trap of Section \ref{rsd} (this time with $\Omega=5J_g/1000$), we plot the large Fermi surface as a solid red line in Fig. \ref{momentumdistributions}(a).  In Fig. \ref{momentumdistributions}(b) we plot the quasi-momentum distribution $\tilde{n}_{LDA}[\bm{k}]$ for the trapped system alongside the large Fermi surface of the translationally invariant system.

It is worth appreciating that the large Fermi surface actually survives smearing by the trap better than it would for free fermions.  This is in part because in the averaging over the trap only Fermi surfaces between the black and red dotted lines of Fig. \ref{momentumdistributions}(a) contribute (for free fermions the averaging would contain arbitrarily small Fermi surfaces).  Another reason is that, as we approach the edge of the conduction atom cloud and the local Fermi surface recedes toward the line $\left|k_x\right|+\left|k_y\right|=\pi$, the height of the discontinuity also decreases to zero.

\section{Self-Consistent Diagonalization}
\label{secSCD}

In order to probe mass enhancement in the heavy fermion state we would like to study dynamics of the conduction atom cloud in response to a trap displacement.  This type of dynamics has been shown to be an excellent diagnostic tool for strongly correlated systems in experiments using bosonic as well as fermionic atoms \cite{Fertig:2005p4167,Strohmaier2007,McKay2008}.  At mean-field level this requires us to know the self-consistent ground state; we must move beyond the LDA and self-consistently diagonalize $\mathcal{H}_{\mathrm{M}}$.  Analytic progress is made difficult by the necessity to maintain site dependent hybridizations, and we proceed numerically.  We start by making an initial guess for the $\tilde{V}_i$ and $\mu_{ie}$ using LDA.  We then diagonalize $\mathcal{H}_{\mathrm{M}}$ numerically, calculate new $\tilde{V}_i$ from Eq. (\ref{SelfConsistency}), and repeat until the $\tilde{V}_i$ converge to a self consistent value.  This process is complicated by the need to repeatedly determine the $\mu_{ie}$ that satisfy the constraint of one $e$ atom per site on average (these will change every time the $\tilde{V}_i$ are updated).  A discussion of the procedure we use to determine the correct $\mu_{ie}$ can be found in the Appendix.

To simplify the numerics, we consider a
geometry where the potential changes only in the \emph{longitudinal}
direction (length $aL_l$ and $g$ atom hopping $J_l$), while the \emph{transverse} direction (length $aL_t$ and $g$ atom hopping $J_t$) is homogenous with
periodic boundary conditions.  The motivation for this unusual geometry is to retain a number of constraint equations equal to the longitudinal length, thereby accessing relatively large system sizes (up to $\sim1000$ sites is reasonable).  Adopting a notation $c^{\dagger}_{mn\alpha\sigma}$ for the creation operators, where $m$ labels the site in the longitudinal direction and $n$ labels the site in the transverse direction, we define partially Fourier decomposed
operators
\begin{equation}
c^{\dagger}_{mk\alpha\sigma}=\frac{1}{\sqrt{L_t}}\sum_{n}e^{-ikn}c^{\dagger}_{mn\alpha\sigma}.
\end{equation}
Rewriting the nonconstant part of the mea- field Hamiltonian [Eq. (\ref{HMFT})] in terms of these new operators, it decouples into the sum of $L_t$ effectively 1D
Hamiltonians, which we label by the transverse quasi-momentum $k\in\left\{\frac{2\pi}{L_ta},...,\frac{2\pi}{a}\right\}$:
\begin{eqnarray}
\mathcal{H}_{\mathrm{M}}&=&\sum_{k}\mathcal{H}_k \nonumber \\
\mathcal{H}_k&=&-J_l\sum_{\left<i,j\right>\sigma}c^{\dagger}_{ikg\sigma}c^{}_{jkg\sigma}+V_{\mathrm{ex}}\sum_{i\sigma}\tilde{V}_i\left(c^{\dagger}_{ikg\sigma}c^{}_{ike\sigma}\right. \nonumber \\
&+&\left.c^{\dagger}_{ike\sigma}c^{}_{ikg\sigma}-\tilde{V}_i/2\right)+\sum_{i\sigma}\left(\mu_{ie}-1\right)c^{\dagger}_{ike\sigma}c^{}_{ike\sigma}\nonumber \\
&+&\sum_{i\sigma}\left(-2J_t\cos (ka)+\Omega i^2\right)c^{\dagger}_{ikg\sigma}c^{}_{ikg\sigma}.
\end{eqnarray}
One can give a simple physical interpretation to these new Hamiltonians: the energy gained by a $g$ atom dispersing with quasi-momentum $k$ in the transverse direction $[2J_t\cos (ka)]$ has been incorporated as a chemical potential in a 1D model describing hopping in the longitudinal direction.  The eigenstates are created by quasiparticle operators
\begin{equation}
\label{mcs}
\alpha^{\dagger}_{qk\sigma}=\sum_{i}\left(u^{i}_{qk}c^{\dagger}_{ikg\sigma}+v^{i}_{qk}c^{\dagger}_{ike\sigma}\right).
\end{equation}
Similar to Eq. (\ref{BMCs}), here $q\in\{1,2,...\}$ is an index labeling the eigenstates in order of increasing energy (for a given $k$).  Unless the transverse bandwidth is smaller than the level spacing of the harmonic potential ($4J_t<2\sqrt{\Omega J_l}$), the different $k$ modes cannot be expected to have the same occupation numbers, which leads us to define $k$-dependent Fermi levels $q_F^k$.  For a given $k$, $q_F^k$ is the smallest $q$ for which the mode labeled by $q$ and $k$ is unoccupied.  The ground state is
\begin{equation}
\left|\Psi\right>=\prod_{k\sigma}\prod_{q<q_F^k}\alpha^{\dagger}_{qk\sigma}\left|0\right>,
\end{equation}
and the expectation value in Eq. (\ref{SelfConsistency}) becomes
\begin{equation}
\label{SCDV}
\tilde{V}_i=\sum_{k}\sum_{q<q_F^k}(u^i_{qk}\bar{v}^i_{qk}+\bar{u}^i_{qk}v^i_{qk}).
\end{equation}
\section{Dynamics}
\label{secDynamics}

\subsection{Single particle dynamics in a trap}

Before introducing dynamics in the MFT, we briefly mention some important details of single-particle dynamics in a lattice plus harmonic potential \cite{Rey:2005p141,polkovnikov,hooley,rigol}.  If there are no interactions, the trapping potential is parabolic with curvature $\Omega$, and we work in 1D for simplicity, then the $g$ atoms are described by the Hamiltonian 
\begin{equation}
\mathcal{H}_0=-J_g\sum_{\left<i,j\right>,\sigma}c^{\dagger}_{ig\sigma}c^{}_{jg\sigma}+\Omega\sum_{i\sigma}i^2c^{\dagger}_{ig\sigma}c^{}_{ig\sigma}.
\end{equation}

A single particle eigenstate $\left|\psi^n_{\sigma}\right>=\sum_i\psi_i^nc^{\dagger}_{ig\sigma}\left|0\right>$ has coefficients $\psi^n_i$ given by the
Fourier components of periodic Mathieu functions,
with the corresponding eigenvalue $E_n$ determined by the function's characteristic
parameter \cite{Rey:2005p141}.  The solutions are complicated, but in certain limits they have a very simple form.  We
will exclusively consider the limit where $q\equiv4J_g/\Omega\gg 1$.  Then $1/\sqrt{q}$ is a
natural small parameter for expanding both the eigenfunctions and
eigenvalues, and to lowest non-trivial order one obtains:
\begin{eqnarray}
\label{assumptions}
\psi^n_i&\sim&\sqrt{\frac{\sqrt{2}}{2^nn!\left(q\pi^2\right)^{1/4}}}\exp\left(-\frac{\xi^2}{2}\right)H_n\left(\xi\right)\nonumber\\
E^n&\sim&\frac{\Omega}{4}\left(-2q+4\sqrt{q}\left[n+\frac{1}{2}\right]-\frac{\left(2n+1\right)^2+1}{8}\right),
\end{eqnarray}
where $\xi\equiv
i\left(4/q\right)^{1/4}$ and $H_n$ are Hermite polynomials.  In this approximation the expectation
value for the center of mass (COM) of the ground state,
evolving due to a small displacement $\delta$ (measured in
lattice units), can be obtained in closed form (the motion of excited states is similar but slightly complicated by various combinatorial factors):
\begin{eqnarray}
\label{COM}
\left<X\left(t\right)\right>&=&\delta\exp\left[-\frac{\delta^2}{x^2}\sin\left(\frac{\Omega t}{8\hbar}\right)^2\right]\times \nonumber \\
&&\cos\left[\left(\omega^*-\frac{\Omega}{4\hbar}\right)t-\frac{\delta^2}{2x^2}\sin\left(\frac{\Omega t}{4\hbar}\right)\right].
\end{eqnarray}
In the above $x\equiv\left(q/4\right)^{1/4}$ is a characteristic
oscillator length (in lattice units) and $\omega^*\equiv\Omega\sqrt{q}/\hbar$ is the
characteristic frequency.  It is interesting to note that there is periodic (non-dissipative) damping of the oscillation amplitude.  In the absence of the lattice this damping would be forbidden by the generalized Kohn theorem \cite{kohn}, according to which COM oscillations in a harmonic potential should be undamped under fairly general conditions.

It is important for our purpose in what follows to observe that the tunneling matrix element is inversely proportional to the lattice-effective mass $m^*=(2J_ga^2/\hbar^2)^{-1}$.  Therefore, the effective oscillator period (at which, for sufficiently large $q$, all low lying single particle modes oscillate) is related to the mass as $\tau^*=\frac{2\pi}{\omega^*}\sim\sqrt{m^*}$.  In what follows, we take the periodicity ($\tilde{\tau}$) of oscillations to define the effective mass for the interacting system ($\tilde{m}$): $\tilde{m}/m^*=\left(\tilde{\tau}/\tau^*\right)^2$.

\subsection{MFT dynamics}

Once we have solved for the MFT groundstate, calculating dynamics is relatively
straightforward.  One can easily write down a time dependent Schr\"{o}dinger
equation for the mode coefficients.  Just as
$\psi_q\left(x\right)\rightarrow\psi_q\left(x,t\right)$ in continuum
quantum mechanics, here we have a discrete analogy $u_{qk}^{j}\rightarrow
u_{qk}^{j}\left(t\right)$, and likewise for the $v_{qk}^{j}$.  The time
dependent discrete Schr\"{o}dinger Eq. reads
\begin{eqnarray}
\label{MFTDmode}
-i\hbar\frac{d u^j_{qk}}{dt}&=&J_l\left(u^{j-1}_{qk}+u^{j+1}_{qk}\right)-\Omega \left(j-\delta\right)^2u^j_{qk} \nonumber \\
&+&2J_t\left[\cos (ka)\right] u^j_{qk}-V_{ex}\tilde{V}_jv^j_{qk} \nonumber \\
-i\hbar\frac{d v^j_{qk}}{dt}&=&-\Omega\left(j-\delta\right)^2v^j_{qk}-\mu_{je}v^j_{qk}-V_{ex}\tilde{V}_ju^j_{qk}.
\end{eqnarray}
Eq (\ref{MFTDmode}) can be obtained formally by considering the coefficients
in Eq (\ref{mcs}) to carry the time dependence and then solving the
Heisenberg equations of motion for the quasiparticles
\begin{equation}
\hbar\frac{d}{d t}\alpha^{\dagger}_{qk\sigma}=i\left[\mathcal{H}^{\delta}_{k},\alpha^{\dagger}_{qk\sigma}\right]
\end{equation}
($\mathcal{H}_k^{\delta}$ being a shift of $\mathcal{H}_k$ by $\delta$ lattice sites).

There is one subtlety in the time evolution: although the exact KLM
Hamiltonian has a local $\mathrm{U}(1)$ symmetry to conserve the $e$ atom density distribution, the MFT does not: the occupation of the localized
band will certainly evolve in time, which we cannot allow.  The
solution is to let the chemical potentials be time dependent.  It is easy to check that
the first time derivative of the $e$ atom density does not depend on
the chemical potentials, and hence the constraints $d n_{ie}/d t=0$ cannot
determine the $\mu_{ie}$.  Instead they must be chosen to satisfy the
second order constraints $d^2n_{ie}/d t^2=0$.  It follows directly from
Eqs. (\ref{MFTDmode}) that $d n_{ie}/d t=0$ initially, since all the mode
coefficients can be chosen to be real.  Keeping the second derivative
zero at all times means the first derivative does not change; it will
always be zero and the local constraints will be obeyed.  To find an explicit formula for the chemical potentials we first express the constraints in terms of the mode coefficients:
\begin{equation}
\label{constraint}
\frac{d^2}{dt^2}n_{ie}=2\sum_{k}\sum_{q<q_F^k}\left(\ddot{v}_{qk}^{i}\bar{v}_{qk}^{i}+v_{qk}^{i}\ddot{\bar{v}}_{qk}^{i}+2\dot{v}_{qk}^{i}\dot{\bar{v}}_{qk}^{i}\right).
\end{equation}
Using the  Schr\"{o}dinger equation to evaluate the time derivatives in Eq. (\ref{constraint}), some algebra leads to:
\begin{equation}
\label{mu}
\mu_{ie}=\frac{-2V_{\mathrm{ex}}\Re\left[d_i\right]^2\left(c_i-f_i\right)-J_l\Re\left[d_i\left(a_i+b_i\right)\right]+\Re\left[d_ix_i\right]}{\Re\left[d_i\right]^2-\Im\left[d_i\right]^2},
\end{equation}
where $\Re$ and $\Im$ are the real and imaginary parts respectively.  The latin letters are defined as:
\begin{center}
\renewcommand{\arraystretch}{2}
\begin{tabular}{| c | c |}

\hline
$a_i=\frac{1}{L_t}\sum^{'}\bar{v}^i_{qk}u^{i-1}_{qk}$ & $b_i=\frac{1}{L_t}\sum^{'}\bar{v}^i_{qk}u^{i+1}_{qk}$\\

\hline
$d_i=\frac{1}{L_t}\sum^{'}\bar{v}^i_{qk}u^i_{qk}$ & $f_i=\frac{1}{L_t}\sum^{'}\bar{v}^i_{qk}v^i_{qk}$\\

\hline
$c_i=\frac{1}{L_t}\sum^{'}\bar{u}^i_{qk}u^i_{qk}$ & $x_i=-\frac{2J_t}{L_t}\sum^{'}\bar{v}^i_{qk}u^i_{qk}\cos (ka)$ \\

\hline

\end{tabular}
\end{center}
where $\sum^{'}\equiv\sum_k\sum_{q<q^k_F}$.
\begin{figure}[h]
\centering
\subfiguretopcaptrue
\subfigure[][]{
\includegraphics[width=6.5 cm]{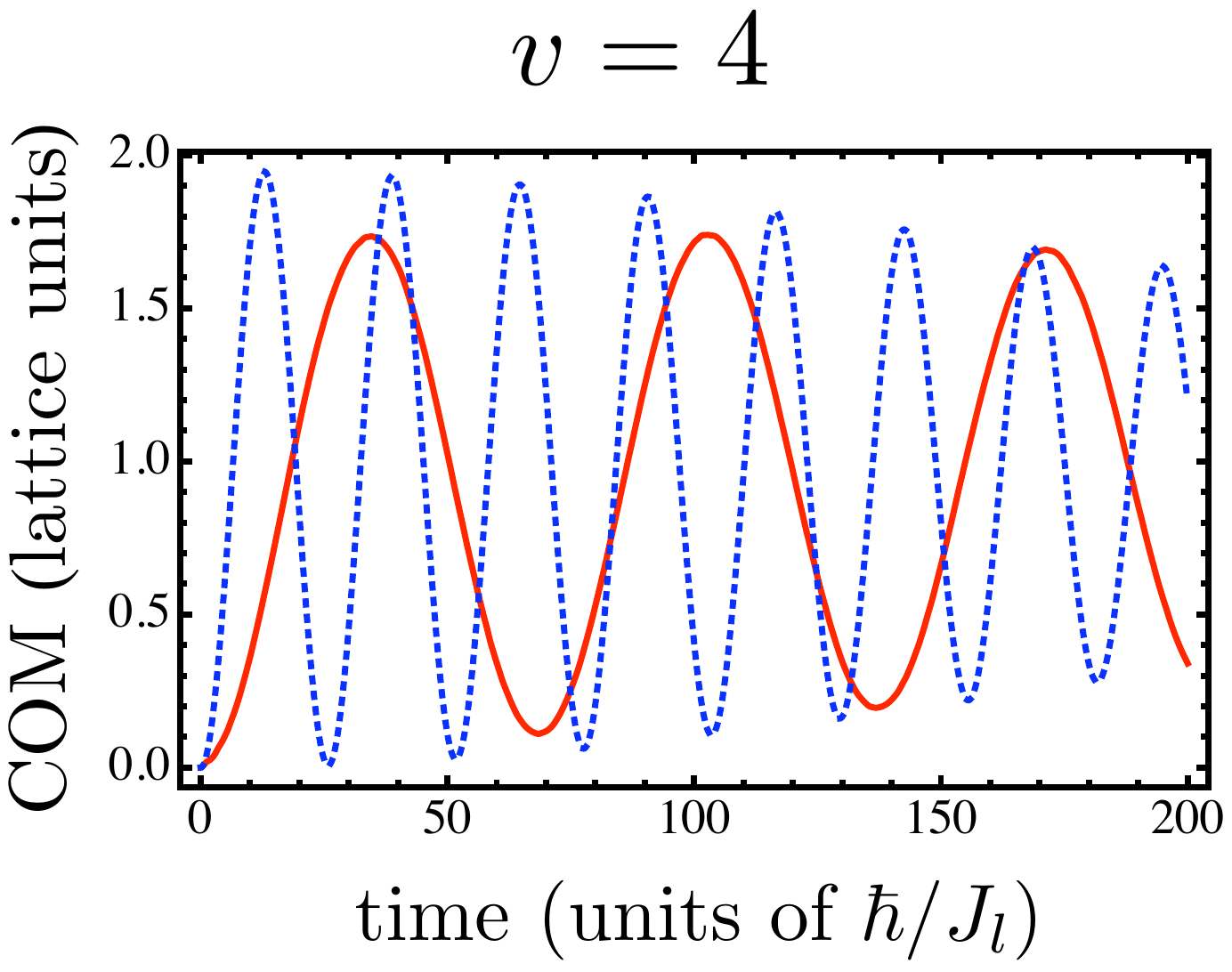}
\label{v4plot}
}
\subfigure[][]{
\includegraphics[width=6.5 cm]{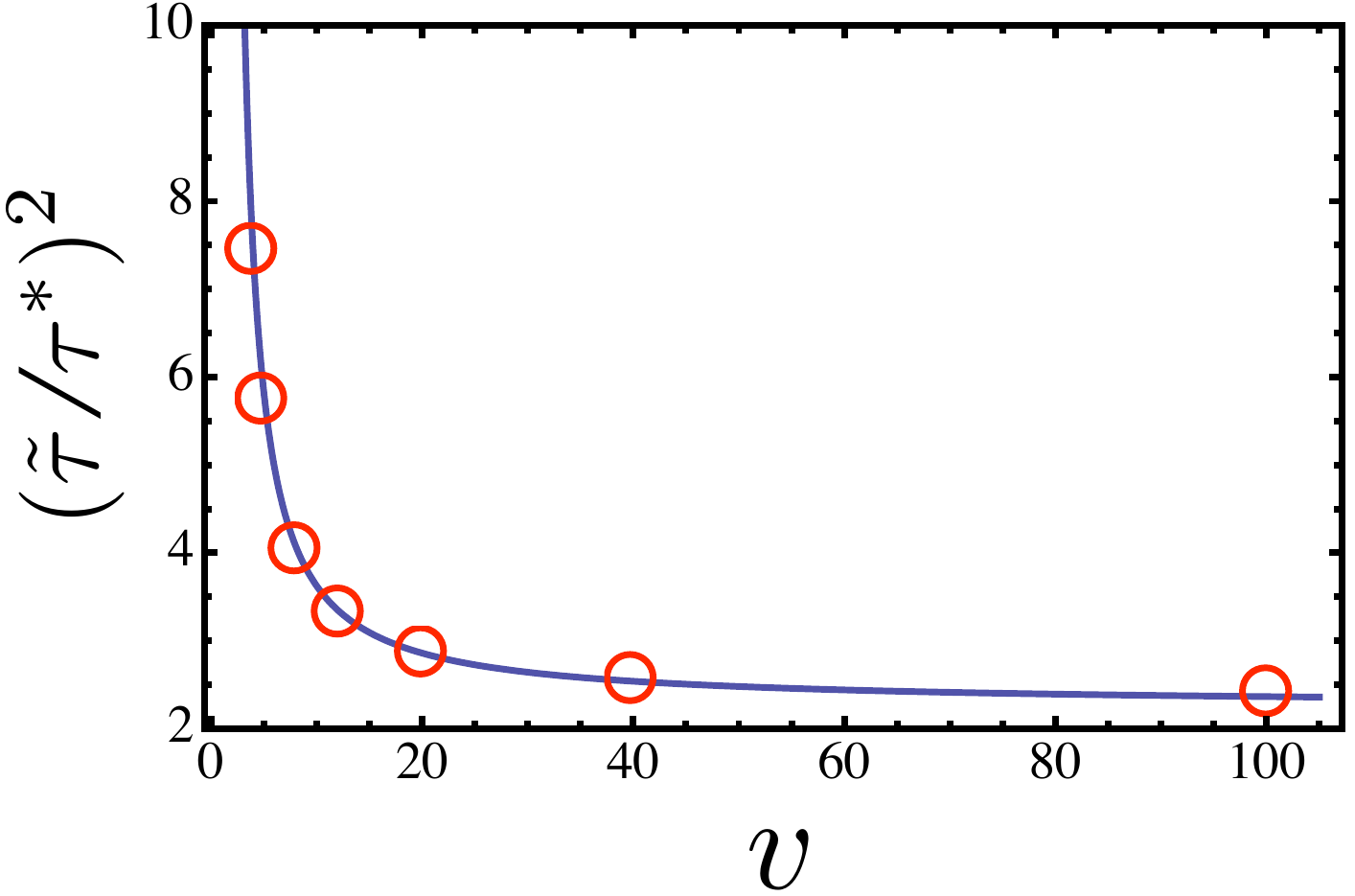}
\label{effectivemassenhancement}
}
\caption{\subref{v4plot} Here we plot the time evolution of the COM after displacement of the trap by one lattice site (red line).  The blue dashed line is the noninteracting dynamics for comparison.  The parameters used in this plot are $q=235$, $v=4$, $L_t=8$, $N_g=40$, and $J_l/J_t=2$.  \subref{effectivemassenhancement}  The effective mass scales like $\tilde{m}/m^*\sim\left(\tilde{\tau}/\tau^*\right)^2$ (see the vertical axis): by extracting the period of oscillation from several traces like the one in \subref{v4plot} at several different values of $v$ we find significant enhancement of the effective mass as $v\leadsto 1$.  The red circles are from self-consistent diagonalization of the MFT, and the blue line is a guide to the eye.}
\label{dynamics}
\end{figure}

\begin{figure}[h]
\subfiguretopcaptrue
\centering
\subfigure[][~$v=1.25$]{
\includegraphics[width=4 cm]{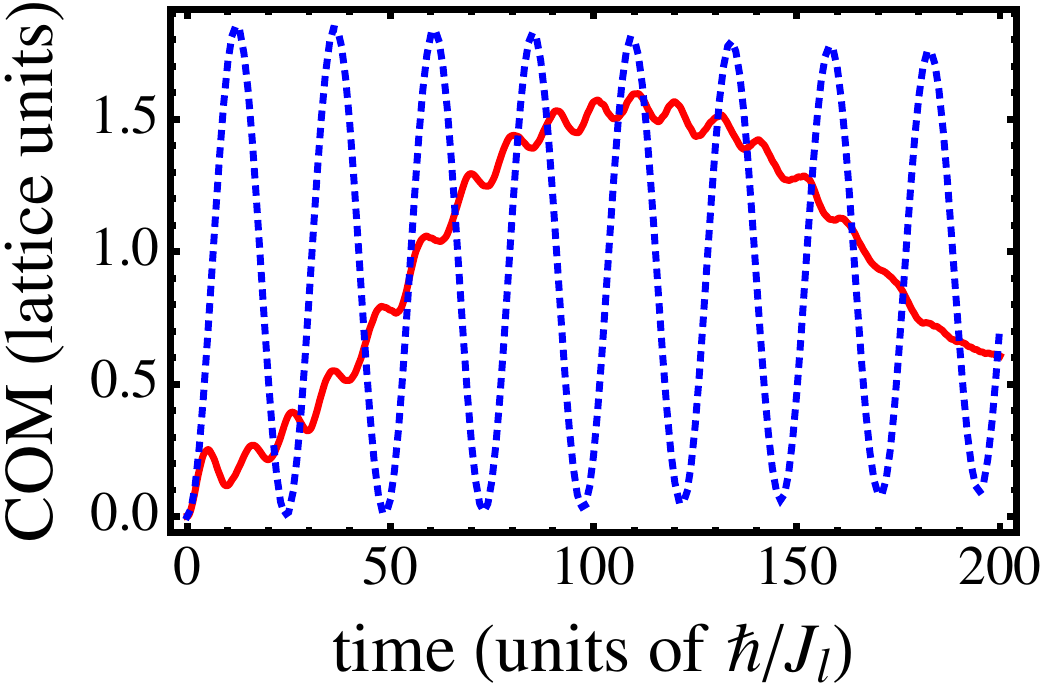}
}
\subfigure[][~$v=1$]{
\includegraphics[width=4 cm]{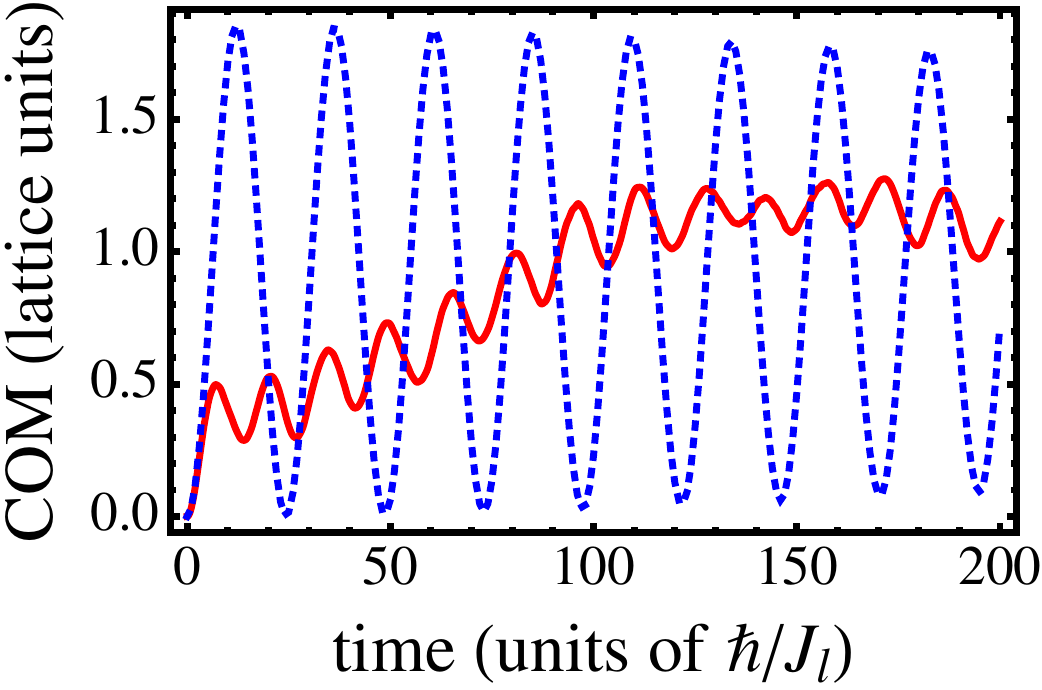}
}
\subfigure[][~$v=0.75$]{
\includegraphics[width=4 cm]{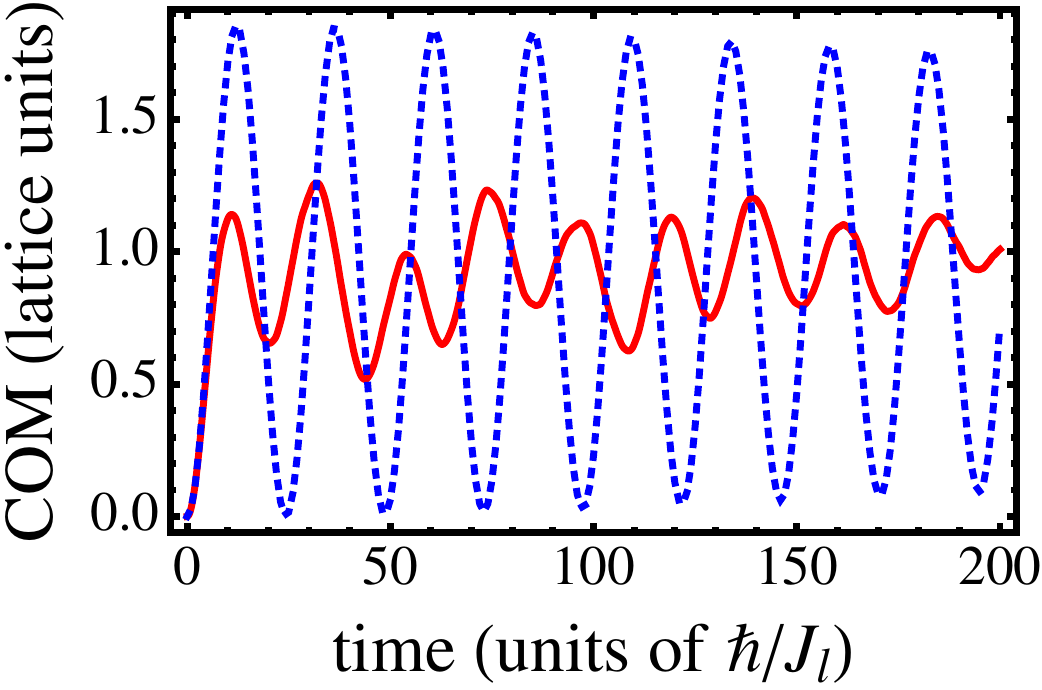}
}
\subfigure[][~$v=0.5625$]{
\includegraphics[width=4 cm]{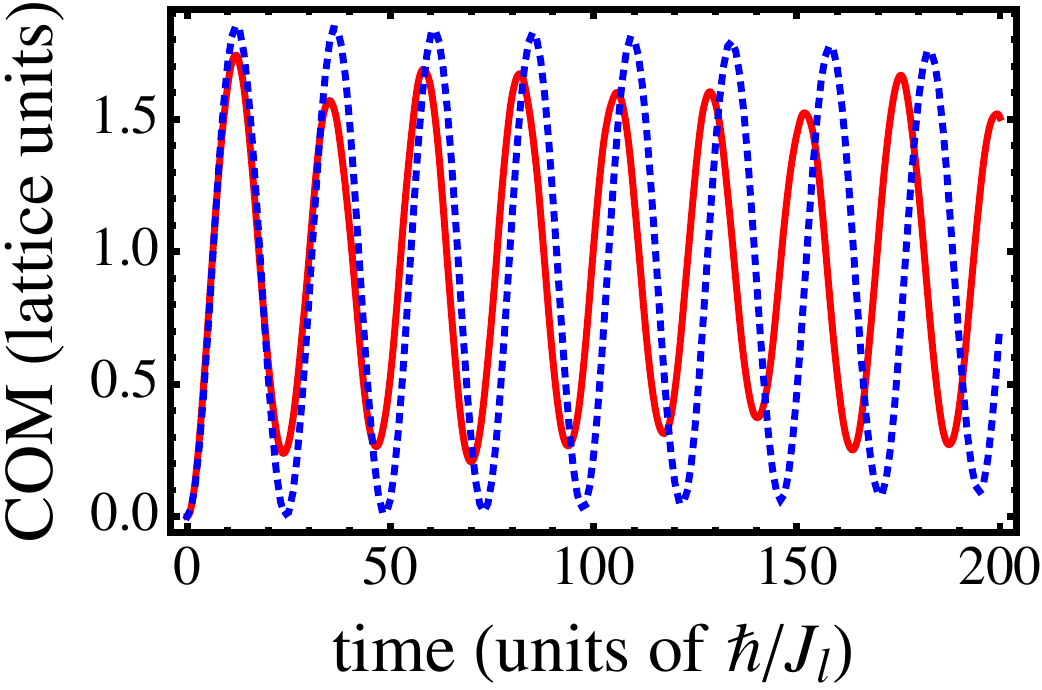}
}
\caption{COM oscillations for decreasing values of $v$, showing how the fast free-particle dynamics emerge on top of the slow quasiparticle dynamics.  The red solid line is from MFT dynamics, and the blue dotted line is the noninteracting solution.  The parameters used in this plot are $q=235$, $L_t=1$, $N_g=6$, and $J_l/J_t=2$.}
\label{gg}
\end{figure}

Using Eq. (\ref{SCDV}) and (\ref{mu}) to write the $\tilde{V}_i$ and $\mu_{ie}$ in terms of the mode coefficients reduces Eq (\ref{MFTDmode}) to a system of $4L_tL_l^2$ coupled first order differential equations, which can be integrated with standard methods.  We calculate the dynamics resulting from displacement of the trap by one lattice site for fixed $J_l,J_t,$ and $\Omega$ at several different values of $V_{\mathrm{ex}}$, and extract the period of COM oscillations (see Fig. \ref{dynamics}).  Since $\tilde{m}/m^*=\left(\tilde{\tau}/\tau^*\right)^2$, Fig. \ref{effectivemassenhancement} demonstrates the emergence of heavy quasiparticles as we decrease $v$.

In the thermodynamic limit of the translationally invariant MFT, the density-of-states effective mass truly diverges as $v\rightarrow0$.  This divergence is not physical, because for sufficiently small $v$ RKKY interactions lead to the development of magnetic order, at which point the paramagnetic MFT is no longer valid \cite{Doniach:1977}.  In the trap, however, the effective mass does not diverge in the small $v$ limit, even within the MFT.  This follows from the discreteness of the spectrum in the non-interacting theory; when the hybridization energy $\tilde{V}V_{\mathrm{ex}}$ becomes smaller than the energy spacing, it cannot significantly change the system properties.  Instead, as we decrease $v$ the relatively fast free particle oscillations emerge on top of the slow quasiparticle oscillations.  Eventually, the dynamics converges to that of the noninteracting system, as shown in Fig. \ref{gg}.

\

\section{Experimental Considerations and Outlook}
\label{secECO}

In order to obtain the heavy fermion metallic groundstate, it is necessary to prepare a Mott insulator of $e$ atoms coexisting with a dilute cloud of $g$ atoms.  To realize the $\mathrm{SU}(2)$ symmetry of the KLM under consideration only two hyperfine states should be initially populated.  In consideration of the lossy nature of $e$-$e$ scattering, the starting point we envision is a Mott insulator of $g$ atoms with doubly occupied sites in the center of the trap.  Such a configuration can be obtained by making the lattice for $g$ atoms sufficiently deep and the parabolic trap sufficiently tight.  The deep lattice for the $e$ atoms is unoccupied but already established.  Taking advantage of the mean-field energy shift associated with double occupied sites, it is in principle possible to make the transfer $\left|gg\right>\rightarrow\left(\left|eg\right>+\left|ge\right>\right)/\sqrt{2}$ with high efficiency.  The sites occupied by a single $g$ atom (at the edge of the trap) can also be addressed independently, transferring the atoms on them into the $e$ lattice.  If the $g$ lattice is adiabatically lowered, we are left with the configuration desired.  For sufficiently large $v$ this procedure can also yield a unit filling plateau at the trap center (the Kondo insulator), but to observe the second layer in the shell structure discussed in Section \ref{rsd} (which has 3 atoms per site in the center of the trap, 1 $e$ and 2 $g$), the trap should be adiabatically tightened until $n_g>1$ at the trap center.

The observation of dipole oscillations over a small number of lattice sites ($\lesssim8$) has precedent in several experiments done with alkali-metal atoms \cite{Fertig:2005p4167,Strohmaier2007}.  Usually the COM velocity of the cloud as a function of time is measured directly by TOF imaging, and the COM position is inferred from simple kinematics.  There are also promising proposals to measure the COM expectation value in a non-destructive and precise manner by coupling the atomic motion to unpumped cavity modes \cite{Peden:2009p5704} (the COM motion is inferred indirectly by its relationship to measurable quadratures of the photon field.)

For the above mentioned experiments to be feasible, it is important that the requisite temperatures are within the reach of current technology.  In real metals heavy fermion behavior develops only below the Kondo temperature $T_K\sim J_ge^{-1/v}$.  For $v\ll1$, which is always the case in heavy fermion metals, the exponential suppression guarantees that $T_K$ is always well below the Fermi temperature of the non-interacting conduction electrons, and that the mass enhancement (which scales as $T_K^{-1}$) is extremely large.  However one should not think of the smallness of $T_K$ as being fundamental to heavy fermion behavior, but simply as the result of the bare parameters ($v\ll1$) which happen to exist in real metals.  A larger Kondo temperature can always be obtained by choosing a larger $v$, at the cost of reducing the mass enhancement.  By artificially creating the KLM in an optical lattice and choosing $v\sim1$ the Kondo temperature could be estimated as $T_K\sim J_g\sim V_{\textrm{ex}}$.  Therefore the temperature need only be smaller than the interaction energy, which is a much better situation than one encounters in proposals to observe super-exchange or RKKY type physics.

\section{Summary}
This paper is motivated by recent progress in cooling and controlling AEMAs \cite{fukuhara1,fukuhara2,stellmer,kraft,escobar,taie,desalvo}, and proposals to use cold AEMAs to implement optical lattice simulations of a broad class of Hamiltonians \cite{Gorshkov:2009p4747,Hermele,cazalilla,fossfeig}, of which the KLM is one example.  Building on these prospects, we make clear several cold-atom experimental signatures which would unambiguously demonstrate the physics found in heavy fermion metals and the closely related Kondo insulators.

Local-density approximation has been used to extend a previously studied mean-field theory to an inhomogeneous setting.  At this level of approximation it is clear that the $g$ atom density distribution holds a smoking gun of the insulating behavior of the KLM at half filling (in the form of a density plateau), and it is plausible that the emergence of a large Fermi surface will be evident despite the inhomogeneity of the trap.  A more careful self-consistent diagonalization of the MFT with a trapping potential in two dimensions is explained in detail, and yields the ground state MFT wavefunction.  We then show how this ground state can be time evolved self consistently upon displacement of the trapping potential, and find that heavy fermion mass enhancement manifests as a slowing of the dipole oscillations when the coupling $v$ is decreased.

\section*{Acknowledgments}
We thank Thomas Gasenzer, Matthias Kronenwett, Alexey Gorshkov, Maria Luisa Chiofalo, Brandon Peden, and Jun Ye for helpful discussions.  AMR and MF are supported by grants from the NSF (PFC and PIF-0904017), the AFOSR, and a grant from the ARO with funding from the DARPA-OLE.  MH is supported by DOE grant number DE-SC0003910, and VG is supported by NSF grant number DMR-0449521.

\appendix*
\section{satisfying the local constraints}
For the initialization of the self-consistent diagonalization procedure, and anytime during the procedure immediately after the $\tilde{V}_i$ have been updated, it is necessary to determine the $\mu_{ie}$ that satisfy the local constraint on the $e$ atom density ($n_{ei}=1$).  Finding the correct $\mu_{ie}$ amounts to finding a root
of the vector-valued function:
\begin{equation}
\label{NLEQ}
\Delta_j\left(\mu_{ie}\right)=1-2\sum_{k}\sum_{q<q_F^k}\left|v^j_{qk}\right|^2,
\end{equation}
since $\Delta_j$ is the deviation from 1 of the number of localized atoms at site $j$.  The dependence of $\Delta$ on the chemical potentials is implicit in the $v^j_{qk}$, because they are obtained by diagonalizing the $\mathcal{H}_{k}$, in which the $\mu_{ie}$ appear
explicitly.  We proceed to solve $\Delta_j\left(\mu_{ie}\right)=0$ via Newton's method, where the gradient is calculated in first order perturbation theory (this is just linear response theory)\cite{Hermele}:
\begin{equation}
\label{susceptability}
\frac{\partial\Delta_j}{\partial\mu_{ie}}=\frac{2}{L_t}\sum_k\sum_{q<q_F^k}\sum_{q'>q_F^k}\Re\left[\frac{\bar{v}^i_{qk}v^i_{q'k}\bar{v}^j_{q'k}v^j_{qk}}{\epsilon_q-\epsilon_q'}\right],
\end{equation}
$\Re$ being the real part.  Newton's method amounts to making successive linear approximations to the nonlinear equation  $\Delta_j\left(\mu_{ie}\right)=0$: we choose a starting point $\mu^0_{ie}$ from the LDA, and then solve
\begin{equation}
\Delta_j\left(\mu^0_{ie}\right)-\sum_i\delta\mu_{ie}\left.\left(\frac{\partial\Delta_j}{\partial\mu_{ie}}\right)\right|_{\mu^0_e}=0
\end{equation}
for the $\delta\mu_{ie}$ by inverting the matrix $\partial\Delta_j/\partial\mu_{ie}$.  We then change $\mu^0_{ie}\rightarrow\mu^0_{ie}+\delta\mu_{ie}$, and repeat until the constraints are satisfied to within a desired tolerance.

This procedure works wherever the hybridizations $\tilde{V}_i$ are finite, but breaks down, for instance, near the edge of the trap.  In general, we initiate the self-consistent diagonalization with a trap strength that is sufficiently weak to allow finite $g$ atom density at the trap edges, and then turn up the trap slowly throughout the iteration.  Whenever the hybridization at the edges begins to vanish, the Fermi level is flanked by two nearly degenerate combinations of a single $e$ atom at the leftmost or rightmost lattice site.  The gradient defined in Eq. (\ref{susceptability}) becomes badly behaved at this point, because the degenerate states split the Fermi level, and they must be excluded from the sum (one can avoid the resulting divergences by exploiting the reflection symmetry about the center of the trap, but inclusion of these degenerate states nevertheless increases the precision necessary to reliably invert the gradient matrix).

\end{document}